\documentclass{article}
\usepackage{amsmath}
\usepackage{amssymb}
\usepackage{amsfonts}

\usepackage{graphicx}

\newcommand{\fsize}{3.5in}

\newtheorem{lem}{Lemma}

\newcommand{\be}{\begin{equation}}
\newcommand{\ee}{\end{equation}}
\newcommand{\bey}{\begin{eqnarray}}
\newcommand{\eey}{\end{eqnarray}}

\newcommand{\pref}[1]{(\ref{#1})}

\newcommand{\al}{a_{\ell}}
\newcommand{\Bl}{B_{\ell}}
\newcommand{\Bn}{B_{\nu}}
\newcommand{\bnj}{\bar{\nu}_g}
\newcommand{\bno}{\bar{\nu}_1}
\newcommand{\bgs}{B_{g,i}^{\bullet} }
\newcommand{\bgsp}{\acute{B}_{g,i}^{\bullet} }

\newcommand{\ca}{{\cal A}}
\newcommand{\cb}{{\cal B}}
\newcommand{\cc}{{\cal C}}
\newcommand{\cd}{{\cal D}}

\newcommand{\co}{{\cal O}}
\newcommand{\cv}{c_v }
\newcommand{\cvi}{c_{v,i} }
\newcommand{\cvj}{c_{v,j} }
\newcommand{\Dn}{D_{\nu}}

\newcommand{\dfcfg}{\delta F_{g,cf} }

\newcommand{\dt}{\Delta t}
\newcommand{\dtc}{\Delta t_{c}}
\newcommand{\dtf}{\Delta t_{f}}
\newcommand{\dx}{\Delta x}
\newcommand{\ei}{e^{(i)}}
\newcommand{\eih}{e^{(i+1/2)}}
\newcommand{\eio}{e^{(i+1)}}
\newcommand{\egi}{\eta_{g,i} }
\newcommand{\fel}{f_{\ell}}
\newcommand{\fgim}{F_{g,i-1/2} }
\newcommand{\fgip}{F_{g,i+1/2} }
\newcommand{\fgipm}{F_{g,i \pm 1/2} }
\newcommand{\fgio}{F_{g,1/2} }
\newcommand{\fgin}{F_{g,N+1/2} }
\newcommand{\fgIm}{F_{g,I-1/2} }
\newcommand{\fgJp}{F_{g,J+1/2} }

\newcommand{\ga}{\alpha }
\newcommand{\gat}{\alpha_{\tau} }
\newcommand{\gb}{\beta }
\newcommand{\gepsi}{\epsilon^{(i)}} 
\newcommand{\gepsih}{\epsilon^{(i+1/2)}} 
\newcommand{\gepsip}{\epsilon^{(i+1)}} 
\newcommand{\ggi}{\gamma_{g,i} }
\newcommand{\gk}{\kappa }
\newcommand{\gkl}{\kappa_{\ell} }
\newcommand{\gklg}{\kappa_{\ell ,g} }

\newcommand{\gll}{\Lambda }
\newcommand{\glll}{\Lambda_{\ell} }

\newcommand{\gs}{\sigma }
\newcommand{\gvep}{\varepsilon}
\newcommand{\kn}{\kappa_{\nu}}
\newcommand{\lo}{\ell_0 }

\newcommand{\ptc}{\Psi{\rm tc}}
\newcommand{\sma}{\sum_{\ell=1}^G \al}
\newcommand{\smg}{\sum_{g=1}^G}
\newcommand{\sml}{\sum_{\ell=1}^G}
\newcommand{\smlp}{\sum_{\ell \neq g}^G}
\newcommand{\suml}{\sum_{\ell}}
\newcommand{\tbi}{{T_i^{\bullet}} }
\newcommand{\tgi}{T_{g,i} }
\newcommand{\tgim}{T_{g-1,i} }
\newcommand{\ugi}{u_{g,i} }
\newcommand{\ugio}{u_{g,i}^0 }
\newcommand{\ugj}{u_{g,j} }
\newcommand{\ugjo}{u_{g,j}^0 }
\newcommand{\ui}{u^{(i)}}
\newcommand{\uih}{u^{(i+1/2)}}
\newcommand{\uio}{u^{(i+1)}}
\newcommand{\ul}{u_{\ell}}
\newcommand{\un}{u_{\nu} }
\newcommand{\vl}{v_{\ell}}

\begin{document}
\pagestyle{headings}

\title{A multigroup diffusion solver using pseudo
       transient continuation for a radiation-
       hydrodynamic code with patch-based AMR\thanks{This
       work was performed under the auspices of the U.S. Department of 
Energy by the University of California Lawrence Livermore National 
Laboratory under contract No. W-7405-Eng-48.}}
\author{Aleksei I.\ Shestakov\\
        Lawrence Livermore National Laboratory\\
        Livermore CA 94550\\
        \and 
        Stella S.\ R.\ Offner\\
        Physics Dept., University of California\\
        Berkeley CA 94720}

\date{}

\setlength{\parskip}{0.5pc}  

\maketitle

\begin{flushright} 
{  \setlength{\baselineskip}{.5\baselineskip} 
\vskip-4.25true in
UCRL-JRNL-224845-REV-4
\vskip3.9true in
}
\end{flushright}

\begin{abstract}
We present a scheme to solve the nonlinear
multigroup radiation diffusion (MGD) equations.  The method
is incorporated into a massively parallel, multidimensional,
Eulerian radiation-hydrodynamic code
with adaptive mesh refinement (AMR).  The patch-based
AMR algorithm refines in both space and time
creating a hierarchy of levels, coarsest to finest.
The physics modules 
are time-advanced using operator splitting.
On each level, separate ``level-solve'' packages
advance the modules.  Our
multigroup level-solve adapts an implicit procedure
which leads to a two-step iterative scheme
that alternates between elliptic solves for each
group with intra-cell group coupling.
For robustness, we introduce pseudo
transient continuation ($\ptc$).  We analyze the
magnitude of the $\ptc$ parameter to ensure
positivity of the resulting linear system, 
diagonal dominance and convergence of the two-step
scheme.  For AMR, a level defines a subdomain for
refinement.   For diffusive
processes such as MGD, the refined level uses
Dirichet boundary data at the coarse-fine
interface and the data is derived from the 
coarse level solution.
After advancing on the fine level, an additional
procedure, the sync-solve (SS), is required in order
to enforce conservation.  The MGD SS reduces to an
elliptic solve on a combined grid for a system of $G$
equations, where $G$ is the number of groups.
We adapt the ``partial temperature''
scheme for the SS; hence, we reuse the infrastructure developed
for scalar equations.  Results are presented.
We consider a multigroup test problem with a known
analytic solution.  We demonstrate utility of $\ptc$
by running with increasingly larger timesteps.  Lastly,
we simulate the sudden release 
of energy $Y$ inside an Al sphere ($r = 15$ cm)
suspended in air at STP\@.  For $Y = 11$ kT, we find
that gray radiation diffusion and MGD produce similar
results. However, if $Y = 1$ MT, the two
packages yield different results.  Our large $Y$
simulation contradicts a long-standing theory
and demonstrates the inadequacy of gray diffusion.
\end{abstract}





\setlength{\parskip}{0.0pc}  
\indent


\section{Introduction}
\label{intro}

This paper describes a numerical method to solve the radiation
multigroup diffusion (MGD) equations.  Two themes are presented.
One is the scheme itself.  We add Pseudo Transient Continuation
$(\ptc)$ to the familiar ``fully implicit'' method of Axelrod
et al \cite{AxDuRh}.  The second theme is code-specific.  Our
MGD solver is embedded in a multidimensional, massively parallel,
Eulerian radiation-hydrodynamic code, which has
patch-based, time-and-space Adaptive
Mesh Refinement (AMR) capability.   Our code's AMR framework
stems from the Berger and Oliger idea \cite{BeOl}
developed for hyperbolic, compressible hydrodynamic schemes.
The idea was expanded by Almgren et al \cite{Alm}
and applied to the type of elliptic solvers required for
the incompressible equations of Navier-Stokes.  Howell and Greenough \cite{HowGre}
applied the Almgren et al framework to the scalar, parabolic
``gray'' radiation diffusion equation, 
thereby creating the start of our radiation-hydrodynamic code.

The AMR framework works as follows.
A domain, referred to as the ``coarse'' or L0 level, is discretized
using a uniform, coarse spatial mesh size 
$h_c$.\footnote{In multiple dimensions, coordinates have their
own mesh spacing.}  After advancing with a
timestep $\dtc$, the result is scanned
for possible improvement.  One may refine
subregions containing a chosen material, at
material interface(s), or at shocks, etc.
Whatever refinement criteria are used,
after the subdomains are identified, specific routines define a
collection of ``patches,'' which cover the subdomains.  In two
dimensions, the patches are unions of rectangles; in 3D, they
are unions of hexahedra.  The patches need not be connected, but
they must be contained within the coarse level.  The patches
denote the ``fine'' or L1 level and are discretized with a uniform,
spatial mesh size $h_f$.  A typical refinement ratio $h_c/h_f$
equals two, but higher multiples of two are also allowed.  

Because the original framework was designed for temporally explicit
hyperbolic schemes, $\dtc$ is restricted by a CFL condition.
This implies a similar restriction for the L1 level timestep $\dtf$.
For the case, $h_c/h_f = 2$, level L1
time-advances twice using $\dtf = \dtc / 2$.
Boundary conditions for level L1 are supplied as follows.
Wherever level L1 extends to the physical boundary,
the level uses the conditions prescribed by the problem.
Portions of level L1's boundary which lie inside the
physical domain have conditions prescribed by time and
space interpolated data obtained from the L0 solution.
For diffusion equations, these conditions are of Dirichlet type.
The numerical solution consists of both coarse and 
fine grid results.  Unfortunately, as it stands,
the composite solution
does not guarantee conservative fluxes across the level boundaries.
To maintain conservation, a separate procedure, dubbed a
sync-solve (SS) is required.  The SS reduces to an elliptic
unstructured grid solve on the composite grid of L0 and L1
levels.  The AMR procedure may be recursive.  That is,
a level L1 grid may generate its own subdomain for
refinement, i.e., a level L2.  In that case, one SS couples results
from levels L1 and L2.  Once the levels advance
to the L0 level time, a SS coupling all three levels
ensues.  For the multigroup equations, the SS 
requires an unstructured grid solve for a coupled system
of reaction-diffusion equations.  Our scheme for a
multigroup SS is an important theme of this paper.

The MGD equations stem from a discretization of the
multifrequency radiation diffusion equations.
The latter is an approximation to the equations of
radiation transfer, obtained by assuming the
matter to be optically thick, which suppresses the
directional dependence of the radiation intensity.
Details of the derivation may be found in various sources:
Mihalas and Mihalas \cite{MM2},
Zel'dovich and Raizer \cite{ZelRai},
Pomraning \cite{Pom}.

The gray radiation diffusion equation is a simplification of the
MGD equations.  It is essentially a one-group equation and is
derived by integrating over all frequencies.  Surprisingly,
it gives very good results in many cases.  However, it clearly
cannot display frequency-dependent effects.  When those are important,
it gives incorrect results.  Unfortunately, unless one solves a
problem with both gray and MGD, one never knows when the former
is adequate.  

We now summarize the paper.
Our MGD scheme consists of two parts.
Sections~\ref{levelsolve} and \ref{MGanlsys} develop the
level-solve algorithm, which is applied on each level.
Section~\ref{levelsolve} develops the equations, the
discretization, and our $\ptc$ scheme.
Section~\ref{MGanlsys} proves three lemmas which determine the
initial magnitude of the $\ptc$ parameter $\gs$.  
Our philosophy for $\gs$
is as follows.  The result of the level solve is the
time-advanced radiation group energy density, which physics dictates 
to be nonnegative.  Zeroing anomalously negative values
is not an option since they are the correct conservative solution to
the linear system that stems from the discretization of the
system.  Thus, the unphysical result nonetheless 
conserves energy.  The difficulty is avoided if in the
original formulation of the linear system $ A x = b$, 
$A$ is an M-matrix and the right-hand-side (RS) is nonnegative.
Since we solve $ A x = b$ using an iterative scheme, the
magnitude of $\gs$ is determined to ensure $b \ge 0$, a
diagonally dominant $A$, and that the iterations converge.
To a large extent, we are guided by Pert~\cite{Pert}, who
discusses how and why the solution to a discretization
of an equation may be unacceptable from a physical standpoint. 
For a first reading,
section~\ref{MGanlsys} may be skipped;
the analysis of the required magnitude of $\gs$ is not
needed for the subsequent sections.

We note that $\ptc$ is widely used to solve nonlinear
systems of equations.  It is closely related
to the Inexact Newton Backtracking Method by
Shahid et al \cite{ShTuWa}.  When applying $\ptc$ to a Newton
solver, the basic idea is to limit the change to the iterates
when one is far from the root but not restrict the change
as one approaches the root.  With $\ptc$, limiting is done
by the magnitude of the pseudo-timestep.  Kelley and Keyes
\cite{KelKey} put $\ptc$ on a solid analytic framework
by examining the three regimes of $\ptc$:
small, medium, and large pseudo-timesteps.  In the last regime,
$\ptc$ recovers Newton's second order of convergence.

Our $\ptc$ implementation differs from the norm.
Standard applications typically detect when a problem is 
``hard'' and then reduce the timestep or some other parameter 
by an arbitrary amount.
However, this method will not work for us because 
our solver is embedded in a time-dependent multiphysics code
with separate modules for compressible gasdynamics,
heat conduction, radiation transport.  Our MGD solver is called
numerous times during the course of a simulation.  (If running,
with AMR, multiple times per physical time advance.)
Although the physical $\dt$ is controlled by various
means, and depending on the problem can vary many orders of
magnitude, we require a MGD solver that works under all conditions.
Our $\ptc$ approach is similar to the one of Shestakov et al \cite{ShHoGr}.
We set the initial magnitude of the $\ptc$ parameter to ensure
that for the first step, our iteration scheme
converges and that the result is physical.
We note that our usage of $\ptc$ is nearly equivalent to having
the MGD module time-advance not in a single (physical) step $\dt$, but in
smaller time increments until the desired time $t^0 + \dt$ is reached.
Some colleagues refer to the process as ``sub-cycling'' the radiation module.
It is easy to show that the lemmas of Sec.~\ref{MGanlsys} still apply
for sub-cycling.

Section~\ref{mgamr} describes the second part of our solver,
viz., the sync-solve.  
Section~\ref{rapAMR} contains results.
Three problems are presented.  The first, in Sec.~\ref{linwin},
displays the accuracy
of the method and its convergence properties: first order in time
and second order in space.  Section~\ref{PTCrobust} demonstrates
the utility afforded by $\ptc$.  For hard problems, it
accelerates convergence; for very hard problems, $\ptc$ is
indispensable.  Section~\ref{hotball} models the
explosive expansion of a hot metal sphere suspended in cold air.
The simulation couples all of the code's physics modules.
The problem is an ideal candidate for AMR since effects propagate
a large distance away from the source, yet in early times,
resolution is needed only near the sphere.  The problem
also demonstrates the necessity of multigroup diffusion.
We find 
that if the sphere's energy is very high, gray diffusion 
gives the wrong answer.  For a 1 MT energy source,
our MGD simulation contradicts results of Brode \cite{Bro},
who used gray diffusion. 
Section~\ref{conclusion} contains concluding remarks.

There are three appendices.  Appendix~\ref{table}
gives a table of exact values for 
the test problem described in section~\ref{linwin}.
Appendix~\ref{apb} discusses situations that
may complicate attaining a diagonally dominant matrix
when discretizing the multigroup system.
Appendix~\ref{apc} presents a spatial convergence
analysis of the multigroup system when
running in ``production'' mode, that is, with a
dominant flux limiter and with AMR.

\section{Level Solve}
\label{levelsolve}

Ignoring velocity terms and Compton scattering, 
the multifrequency radiation equations (CGS units)
(Mihalas and Mihalas \cite{MM2}) are:
\bey
  \partial_t \un 
      & = &  \nabla \cdot \Dn \, \nabla \un +
         c \, \rho \, \kn \, ( \Bn - \un  ) \, , \label{ueq2} \\
  \rho \, \partial_t e & = &  
        - c \, \rho \int_0^{\infty}  d \nu \,
         \kn \, ( \Bn - \un  )  \; , \label{Teq2} 
\eey
In \pref{ueq2}--\pref{Teq2}, 
$\un$ and $e$ 
represent the spectral radiation energy density and matter 
specific energy, respectively.  The former is a function of
position $x$, time $t$ and frequency $\nu$, while $e$
is a function of the mass density $\rho$ and material
temperature $T$, quantities which themselves depend on
$x$ and $t$.  Evolution of $\rho$ is governed by
hydrodynamics.  Hence, in our context, $\rho$ is
a known function.  Introducing the specific heat
$\cv = \partial e / \partial T$ turns \pref{Teq2}
into an evolutionary equation for $T$; hence, the
left-hand-side (LS) becomes $\rho \cv \partial_t T$.  
The subscript $\nu$
designates that the term varies with
frequency.  In \pref{ueq2}--\pref{Teq2},
$c$ denotes the speed of light,
$\kn$ the absorption opacity, and $\Bn$ the Planck function,
\[
  \Bn = (8 \pi \, h / c^3) \, \nu^3 / \,[\exp(h \nu / k T) - 1] \;\;\;\; 
  ( {\rm erg}\;{\rm sec}\;{\rm cm}^{-3}) \, ,
\]
where $h$ and $k$ are the Planck and Boltzmann constants,
respectively.
The diffusion coefficient $\Dn$ depends on the total inverse
mean free path $\chi_{\nu} = \rho \kn + \rho \kappa_{\nu,s}$,
where $\kn$ and $\kappa_{\nu,s}$ are the absorption
and scattering opacities, respectively.  (The
opacities are also functions of material composition, $\rho$
and $T$.)  In \pref{ueq2}, the term $- \Dn \nabla u$ denotes
the spectral radiation energy flux.
To limit energy
streaming faster than $c$, a flux limiter
is introduced, e.g.,
\be
  \Dn = c \, / \, [\,3 \chi_{\nu} + | \nabla(\un) | / \un ] \, . \label{difcof}
\ee

The multigroup equations are derived as follows. 
The frequency domain is discretized
into $G$ {\em groups\/} with boundaries
$\{ \nu_g \}_{g = 0}^G$ satisfying
\[
  0 \le \nu_0 < \nu_1 < \ldots < \nu_G < \infty \, .
\]
The equations are integrated over groups.  We define
\[
  u_g(\, x, \, t) = \int_g \un =
  \int_{\nu_{g-1}}^{\nu_g} d \nu \, \un \, .
\]
Time derivatives are replaced by differences
and the system is multiplied by the timestep $\dt$.
Integration of the transport and absorption terms
requires defining group-averaged opacities.  Linearizing
the Planck function about a known temperature $T^*$, the
absorption term is expressed as
\[
  \int_g \kn \, ( \Bn - \un  ) = 
    \gk_g \, [ \, B_g + B_g' ( T - T^*) - u_g  \, ] \, ,
\]
where $\gk_g$ is the group-averaged absorption opacity,
$B_g = \int_g \Bn|_{T = T^*} $, and
$B_g' =  \int_g (\partial \Bn / \partial T)|_{T = T^*} $.
In a semi-implicit scheme, $T^* = T^0$, where $T^0$ is the
temperature at the start of the time cycle.  
For fully implicit differencing, we must iterate until $T^*$
converges to $T$.  For the transport term, we define
\[
    \dt \int_g \nabla \cdot \Dn \, \nabla u = 
    \nabla \cdot D_g  \nabla u_g \, ,
\]
where $D_g$ depends on a group-averaged inverse
mean free path $\chi_g$.  Note that $\dt$ has been
absorbed into $D_g$.

The above definitions yield the multigroup equations,
\bey
 0 & = & u_g^0 -  u_g + \nabla \cdot D_g  \nabla u_g
    +  K_g( \, u, \, T) \, , \;\;\;
    g = 1, \, \ldots , \, G \label{ueq4} \\
 0 & = & \rho \, \cv ( T^0 - T ) - 
   \sml K_{\ell}( \, u, \, T)  \; , \label{Teq4}    
\eey
where $u_g^0$ and $T^0$ denote values at the start of the time-advance, 
\begin{eqnarray}
  K_g( \, u, \, T) & = & a_g \, 
     [ \, B_g + B_g' \, ( \,T - T^* \,)- u_g \, ]  \; , \nonumber \\
  a_g      & = & \dt \, c \, \rho \, \gk_g \; .  \nonumber
\end{eqnarray}

Equations~\pref{ueq4}--\pref{Teq4} comprise a nonlinear system
with the strongest nonlinearity due to the emission term $B$.
To a lesser extent, opacities 
also have a temperature
dependence and for nonideal gases, so does $\cv$.
However, for ease of solution,
we may choose to view \pref{ueq4}--\pref{Teq4}
as a linear system in which case all coefficients
are evaluated at the old temperature $T^0$. 
For simulations in which matter and 
radiation are tightly coupled, i.e., where we expect to
have $\un = \Bn$, the
solution to the semi-implicit difference equations
is $u_g = B_g + B_g' \, ( T - T^0)$, with $B_g$ and
$B_g'$ evaluated at $T = T^0$.  For
high frequencies,
$\lim_{\nu \rightarrow \infty} (\Bn / \Bn') \sim 1/ \nu$;
hence, $B_g \ll B_g'$ for large $g$.  Unfortunately, if the 
temperature is decreasing, i.e., if $( T - T^0) < 0$,
the linearized emission term is negative for large $g$,
leading to the unphysical result: $u_g < 0$. 
On the other hand, if we are able to iterate on $T^*$
so that it converges to $T$, then in tightly coupled
simulations, we obtain the desired solution
$u_g = B_g$ with $B_g$ evaluated at the advanced temperature.

In our code we provide both options, i.e., solving a
linear system, or converging on the implicit
source.\footnote{At the time of this writing, opacities
and $\cv$ were time-lagged.}
In either case, solving \pref{ueq4}--\pref{Teq4}
on a large domain with many groups presents a formidable
task.  To facilitate the task, we introduce pseudo transient
continuation ($\ptc$) and replace the zeros on the LS of
\pref{ueq4}--\pref{Teq4} with the $\ptc$ derivatives,
\[
  \tau \, ( u_g - u_g^*) \;\;\; {\rm and} \;\;\;
  \rho \, \cv \, \tau \, ( T - T^*) \; ,
\]
where $\tau \ge 0$, the inverse of the
pseudo-timestep, is the $\ptc$ parameter whose magnitude
is at our disposal. 

The variables $u_g^*$ and $T^*$  represent advances in pseudo time;
they always appear on the LS of \pref{ueq4}--\pref{Teq4}.
As mentioned above, we provide the option of running in either
semi-implicit (SI) or fully-implicit (FI) mode.  With SI,
since $B_g$ is linearized about $T = T^0$, in the definition of
the coupling term $K_g$, we substitute $T^0$ for $T^*$.
However for FI, $K_g$ is defined as above;
$B_g$ is linearized about the pseudo time temperature $T^*$.
The two modes lead to subtle differences
in the scheme, as shown below.

For the FI scheme, if the matter equation is solved for the temperature
change, we obtain
\bey
 \delta^{-1} \, ( \,T - T^* \, ) = 
     \rho \, \cv \, ( T^0 - T^* ) - 
   \sml \al \,  ( \Bl - \ul ) \, , \label{dTeq}
\eey
where
\be
  \delta^{-1} =  \rho \, \cv \, \gs
    + \sml \al \, \Bl'  \;\;\;\;  {\rm and} 
  \;\;\;\; \gs \doteq 1 + \tau \, .\label{deldef} 
\ee
The domain of relevance $\tau \ge 0$ corresponds to $\gs \ge 1$.

For the SI scheme, the temperature change is,
\be
 \delta^{-1} \, ( \,T - T^0 \, ) = 
     \rho \, \cv \, (\gs-1) \, ( T^* - T^0 ) - 
   \sml \al \,  ( \Bl - \ul ) \, . \label{dTeqSI}
\ee
The term $\delta$ is defined as above, but
$B_g$ and $B_g'$ are evaluated at $T = T^0$.

For the FI scheme, if \pref{dTeq} is substituted into the
equation for $u_g$, we obtain 
\bey
 \lefteqn{ 
   - \nabla \cdot D_g \, \nabla u_g  
   + ( \, \gs + a_g \, ) \, u_g 
   - f_g \sml \al \, \ul = }
   \nonumber \\
  & &  u_g^0 + ( \, \gs -1 \, ) \, u_g^* + a_g \, B_g 
       + f_g  \left( \, \rho \, \cv \, ( T^0 - T^* )
                       - \sml \al \, \Bl
              \right) \, , \label{ulineq}
\eey
where $f_g \doteq \delta \, a_g \, B_g'$.
Equation~\pref{deldef} implies $f_g < 1$, for all $g$.
For the SI scheme, the RS of \pref{ulineq} changes:
$\rho \, \cv \, ( T^0 - T^* )$ is replaced with
$\rho \, \cv \, (\gs-1) \, ( T^* - T^0 )$.

Equation~\pref{ulineq} corresponds to a linear system
\[
  \ca \, u = w
\]
of order $(N \times G)$, where $N$ is
the number of mesh cells and $G$ the number of
groups.  The first term on the LS of \pref{ulineq} consists 
of second order, central differences over space.  We write this term as
\[
  - \nabla \cdot D_g \, \nabla u_g = 
    + \cd_{d,g} \, u_g - \cd_{o,g} \, u_g \, .
\]
The first part represents multiplication of the
vector $u_g$ by a {\em diagonal\/} matrix; the second term
denotes multiplication by the off-diagonal part.
The coefficients of $\cd_{d}$ and $\cd_o$ are nonnegative.

On the LS of \pref{ulineq}, the term
$- f_g \sml \al \, \ul$
is referred to as the ``re-emission source''
\cite{Moretal}, since 
it represents radiation energy absorbed by matter and
re-emitted.  
If we define the {\em column} vectors $f$ and $a$ with components
$f_g$ and $a_g$, respectively,
the re-emission term is expressed as the
matrix-vector product
\be
  - \; ( \, f \, a^{{\rm T}} \, ) \, u \; ,  \label{reterms2}
\ee
where  $a^{{\rm T}} = {\rm transpose} \, ( a )$,
and $u$ is the column vector of unknowns.
Since the re-emission term does not
couple cells, \pref{reterms2} corresponds to
separate products: one per cell, with each product
of order $G$.

These observations allow expressing the matrix as
\be
  \ca = \gll - M_1 - M_2  \label{asplit},
\ee
where $\gll$ is diagonal, $M_1$ contains the offdiagonal
terms due to the (spatial) diffusion term,
and $M_2$ contains  the offdiagonal
terms due to interfrequency coupling.
The corresponding elements are
\bey
   \gll_g & = & \cd_{d,g} + \gs + a_g - f_g \, a_g \, ,
    \nonumber   \\  
  (M_1 \, u)_g & = & \cd_{o,g} \, u_g  \, ,\nonumber   \\ 
  (M_2 \, u)_g & = &  f_g \smlp \al \, \ul \, . \nonumber    
\eey  
The decomposition \pref{asplit} leads to the iterative
scheme proposed by Axelrod et al \cite{AxDuRh}, which improves
a guess $\ui$ by successively solving
\bey
  (\, \gll - M_2 \, ) \, \uih & = & w + M_1 \, \ui \label{iterh} \\
  (\, \gll - M_1 \, ) \, \uio & = & w + M_2 \, \uih \label{itero} \; .
\eey
We solve \pref{iterh}--\pref{itero} until $\ui$ converges.
Convergence is gauged  by evaluating the 1-norms of $w$
and the residual $r = w - \ca \, u$; the latter defined as,
\[
  r = w - \ca \, \uio = M_2 \, ( \uio - \uih ) \, .
\] 
The procedure is fast since multiplication by $M_2$ is
local to each cell, which is very convenient if the
spatial domain is decomposed on multiple processors.

We now review the derivation of the system $\ca \, u = w$.
First, we assume that
$\ptc$ is not used, i.e., that $\gs = 1$ in \pref{deldef}-\pref{ulineq}.
For the SI scheme, the terms
$\Bl$ and $\Bl'$ are evaluated at $T = T^0$.  For FI
differencing, we require two types of iterations.  
Equations~\pref{iterh} and \pref{itero} comprise the {\em inner\/}
iteration.  It is initialized with $u^{(0)}$ equal to $u^0$.
Once the inner iteration has converged
to sufficient accuracy, \pref{dTeq} yields the new temperature.
The SI scheme essentially
ends after the inner iteration converges (see below).
For FI
differencing, after $T$ is computed,
the {\em outer\/} iteration sets $T^* = T$,
recomputes $\Bl$ and $\Bl'$ at $T = T^*$ and returns to the
inner iteration.  The outer iteration halts when $T^*$
converges.

If $\ptc$ is invoked, more care is required because
when $\gs > 1$, the system $\ca \, u = w$ is not a true
discretization of the multigroup equations.  
Despite this complication, $\ptc$ brings
robustness to the scheme.  The $\ptc$ parameter $\tau$
plays the role of an inverse timestep in pseudo-time.
In principle, we could set $\tau$ to a large value and solve
a succession of linear systems.  The solution of each system
represents an advance in pseudo-time.
We continue advancing until we reach the pseudo-time steady-state.
This is easily seen by letting $u_g^* = u_g$ on the RS of
\pref{ulineq} and moving the term to the LS\@.
However, making $\tau$ large is not practical as it 
involves many pseudo-time advances.  Furthermore,
the intermediate pseudo-time results are of no interest.
Consequently, we adopt the strategy of making $\tau$
as small as possible.  We discuss the strategy in 
section~\ref{MGanlsys}.

$\ptc$ may be used with either SI or FI differencing.
In the former, once 
\pref{iterh} and \pref{itero} are converged, \pref{dTeqSI}
yields the new temperature $T$.  We then compute the
1-norm of the ``nonlinear'' residual of the linearized equation
for the matter energy,
\be
   r_{nl} = V  
     \left( \rho \, \cv \, ( T - T^0 ) - 
   \sml \al \,  [ \Bl + \Bl' \,( T - T^0 ) - \ul ] 
     \right) \, , \label{rnl}
\ee
where $V$ is the cell volume.
The residual is compared with the 1-norm of
the matter ``energy'' $ V  \rho \, \cv \, T$, and
in order to monitor stagnation,
 it is also compared with the energy change 
 over the pseudo-timestep
$ V  \rho \, \cv \, (T - T^*)$.
With FI, the temperature
$T$, obtained from \pref{dTeq}, is used to compute the
emission $\Bl$.  The residual $r_{nl}$ is defined
as in \pref{rnl}, except without the $\Bl' \,( T - T^0 )$ term. 

Unfortunately, unless the iterations converge to round-off 
accuracy, energy may not be conserved. 
Lack of conservation stems from values of user-set parameters that
control stopping criteria for the iterations.  For example, it may be
efficient to halt once $||r_{nl}||_1 < 10^{-6}$,
and the norm of the iterates $|| (\Delta T)/T ||_{\infty} < 10^{-2}$ since
continuing brings little noticeable (visual) improvement to the
solution.  However, if one were to stop at that point, energy may not
be conserved to desired accuracy.
To restore conservation,
we provide the option of an additional step.  After
the iterations stop, we assume that the last computed
temperature $T$ is ``frozen'' and use it to compute
emission.  In the SI scheme, emission into the $g$th
group is defined as $S_g = B_g + B_g' ( T - T^0)$,
where $B_g$ and $B_g'$ are evaluated using
$T^0$.  (To prevent unphysical behavior, $S_g$
is not allowed to be negative.)
In the FI scheme, we evaluate $B_g$ using $T$ and
set $S_g = B_g$.  Having a known emission allows us
to compute the energy-conserving radiation field.
The groups decouple.  For $g = 1, \, \ldots, \, G$,
we solve
\[
  - \nabla \cdot D_g \, \nabla u_g  
   + ( \, 1 + a_g \, ) \, u_g = 
   u_g^0 + a_g \, S_g \, .
\]

After computing $u_g$, the matter energy density change is
\[
  \Delta {\cal E} = - \sml \al ( \, B_g - u_g \, ) \, ,
\]
where, if using the SI scheme, $B_g$ is linearized about
$T= T^0$, or with FI, is evaluated at $T$.
The quantity $ V\, \Delta {\cal E}$
represents the average energy change of the matter.
In cells with more than one material, we adapt
a suggestion of Zimmerman~\cite{george},
which simulates intra-cell gray diffusion.
The scheme assumes each material resides in its
own sub-volume.  We solve for separate, frequency-averaged
radiation energy densities and matter temperatures in the
sub-volumes.  The energy change of the materials depends
on the individual, frequency-averaged opacities
as well as on $\Delta {\cal E}$.

We now briefly describe the spatial discretization.
We largely follow procedures described by Howell and Greenough~\cite{HowGre}
(H\&G) and Shestakov et al~\cite{ShHoGr}.
Our MGD solver is embedded in an Eulerian radiation-hydrodynamic
code with cell-centered fundamental variables: $ \rho, \, u_g$, etc.
The code has distinct 1, 2, and 3D executables;  mesh cells
are line intervals, rectangles, and rectangular hexahedra, respectively. 

In 2 and 3D, we discretize the diffusion term $\nabla \cdot D_g \nabla u_g$
using the H\&G subroutines since those are readily available.
We note in passing that H\&G use the
Levermore-Pomraning flux limiter~\cite{LevPom}
instead of the simple expression in \pref{difcof}.  For 1D we have our own
discretization; $\nabla \cdot D_g \nabla u_g$ is written as
\be
  [\, D_{i+1/2} \, (u_{i+1}-u_{i})/h - 
      D_{i-1/2} \, (u_{i}-u_{i-1})/h \, ]\, / \, h \, , \label{divgrd}
\ee
where the group index $g$ is suppressed and where $i$ is the cell index.

The {\em face}-centered
diffusion coefficient $D_{i+1/2}$ is computed as follows.  First,
we modify \pref{difcof} by adding the term $\gb/h$ to the denominator,
where $\gb$ is a small, user-specified constant,
e.g., $10^{-6}$.  After factoring $h$, we obtain
\[
  D = c h \, / \, [\,3 \chi h  + | \Delta(u) | / u + \gb ] \, ,
\]
where we suppress the group index and note that the expression is
to be evaluated on a face.  The denominator is now dimensionless.
The second term is the relative difference of $u$; we discuss
its discretization 
momentarily.  The product $\chi h$ is an optical depth.
In this light, $\gb$ provides a floor to the cell's optical depth.
The aim is to avoid complications with the matrix solve
in case $\chi \rightarrow 0$ and at the same time,
$\un$ is nearly spatially constant, which may easily happen for
high frequency groups.  The face-centered opacity is
an average of the adjoining cell-centered opacities.  
We offer several options.  For the simulations in this paper,
we typically use inverse averaging, but other options
(arithmetric, square root) are also allowed.\footnote{If the
two opacities are very different, inverse averaging:
$\gk_{i+1/2} \doteq 2\gk_{i}\gk_{i+1}/(\gk_{i}+\gk_{i+1}) \rightarrow
2 \min( \gk_{i}, \, \gk_{i+1})$.  Assuming the opacity is
monotone with $T$, the result is nearly the same as what is
commonly done in gray diffusion, viz., forming a face-centered
temperature, $T_{i+1/2} \doteq (T_{i}+T_{i+1})/2$, and 
calculating $\gk_{i+1/2}$ directly with $T_{i+1/2}$.
For example, if $\gk = \gk_0/T^n$
and $T_{i} \gg T_{i+1}$, inverse averaging gives $2\gk_0/T_i^n$
while the face-centered $T$ result is $2^n\gk_0/T_i^n$.
For the free-free gray opacity, $n = 3.5$; hence,
the two results are similar.
Of course, if the opacity is not monotone with $T$, the face-centered
technique is better.  We plan to incorporate that option in the future.
However, we note that multigroup opacities are usually not strong
functions of $T$.}
The term $|\Delta(u) | / u$ is written as
\[
  2 \, |u_{i+1}-u_{i}|/(u_{i+1}+u_{i}) \, .
\]
Other options are also available, e.g., instead of the
arithmetic average, one may substitute
max($\,u_{i+1}, \,u_{i}$) in the denominator.  We plan to extend the above
discretization to higher dimensions.

Cell-centered data, such as $c_v$, are obtained as in \cite{HowGre}.

For coupling to the radiation field in mixed-material cells, we need averaged material
properties, e.g., opacities.  These are obtained by mass averaging.
Suppressing the group index, if $n$ is the material index and
denoting averaging with an overbar, the opacity (cm$^2$/g)
is given by $m \, \bar{\gk} = \sum_n m_n \gk_n$,
where $m_n$ is the mass of the $n^{\rm th}$ material.  Equivalently,
\[
  \bar{\rho} \, \bar{\gk} \doteq \sum_n f_n \rho_n \gk_n \, ,
\]
where $f_n \doteq V_n / V$ is the volume fraction.


  
This concludes the description of the algorithm
used to advance the multigroup equations on an AMR level.
In the following section, we analyze the convergence of
\pref{iterh}-\pref{itero}, and 
we focus on how the $\ptc$ parameter $\gs$
ensures stable, robust iterations, to yield a physical, i.e.,
nonnegative result.

\section{Analysis of $\mathbf{\Psi}$tc}
\label{MGanlsys}

In this section we develop three criteria that
set the
$\ptc$ parameter.  Disinterested readers 
can safely skip the analysis and continue to 
section~\ref{mgamr} where we discuss the implementation of the
multigroup scheme in the context of AMR.

Recalling that $\gs = 1 + \tau$, 
we develop lemmas that set the
{\em initial\/} magnitude of $\tau$,
where by initial we mean the following.
A new value of $\tau$ is determined at each time advance 
for each AMR level. The level advance consists
of nested loops.  For the ``inner'' iterations,
$\tau$ is fixed.  After convergence,
$\tau$  is reset to
$\tau \rightarrow \gat \tau$, where $\gat$ is a user-set
input whose default value is 1/2.  Section~\ref{PTCrobust}
describes an experiment with another setting of $\gat$.
Our strategy for the initial $\tau$ is to ensure
a nonnegative $w$, diagonal dominance, and
a convergent inner iteration.
For the derivation, it is convenient to define
\bey
  \cb   \doteq  \sml \al \, \Bl   
  & , &
  \cb'  \doteq  \sml \al \, \Bl' \; , \label{cBpdef} \\
  C_g' \doteq a_g B_g' / \rho \, \cv \;\; , \;\; 
  \cc \doteq \cb / \rho \, \cv 
  & , &
  \cc' \doteq \cb' / \rho \, \cv \; .  \label{cCdef}
\eey  

\subsection{Positivity of \mbox{\boldmath $w$}}
\label{posw}

Before analyzing the effect of $\ptc$, we examine the scheme's
behavior without it.  
If $\gs = 1$, the term $u_g^*$ disappears from
\pref{ulineq}.
In the following discussion, we
ignore the $T^0 - T^*$ term since
for the SI scheme, or for the first FI inner
iteration, $T^* = T^0$.
Since $u_g^0 \sim B_g$, if either $\dt$ is large
or the coupling is strong, $a_g \, B_g \gg u_g^0$.
Hence, in this case, the RS of the system,
$w \approx a_g \, B_g - f_g \, \cb$,
where $\cb$ is defined in \pref{cBpdef}.
If $\gs = 1$,
$f_g = a_g B_g'/( \rho \, \cv + \cb') = C_g'/(1 + \cc')$.  Hence,
\[
  w \approx a_g \, 
  ( B_g + B_g \, \cc' - B_g' \, \cc )
  \left/
    \left( 1 + \cc'
    \right)
  \right. \, .
\]
Since $\cc$ and $\cc'$ are proportional to $\dt$
times the opacity, 
the sole $B_g$ term in the numerator is swamped by the other two
terms when $\dt$ is large or the matter is optically thick.
In this limit, the sign of $w$ equals
the sign of ($B_g \, \cc' - B_g' \, \cc)$, which
may be negative.

However, with $\ptc$, nonnegativity of $w$ is equivalent to
the inequality
\[
  0 \le p(\gs) =  u_g^* \, \gs^2 + 
    2{\tilde b} \, \gs + {\tilde c} \; ,
\]
where
\bey
  2 \, {\tilde b} & = & 
            u_g^0 - u_g^* + a_g \, B_g 
           + \cc' \, u_g^* \; , \nonumber \\
  {\tilde c} & = & \cc' \, ( u_g^0 - u_g^* + a_g \, B_g \, ) + 
     a_g \, B_g' \,  [ \, T^0 - T^* 
        - \cc \, ] \, , \nonumber
\eey
for the fully-implicit (FI) scheme.
The SI scheme, adds the term $a_g \, B_g' \, (T^* - T^0)$ to the
definition of $2 \, {\tilde b}$.
If $\gs = 1$, we recover the non-$\ptc$ scheme, which
as shown, may have indeterminate sign($w$).
At the end of the section we show that the SI scheme is less
robust.  We first analyze the FI scheme.

For large $\gs$, $p$ is positive---even if $u_g^* = 0$.
The derivative $dp/d \gs$ increases monotonically and
is positive for $\gs = 1$.
If $u_j^* = 0$, $p$ increases linearly with $\gs$ and has
slope $u_g^0 + a_g \, B_g > 0$. 
Hence, we have proved:
\begin{lem}
\label{lem:L1}
If $p|_{\gs = 1} \ge 0$, the {\rm RS} of \pref{ulineq} is nonnegative
for all $\gs \ge 1$.  Otherwise, \newline
(1) If $ u_j^* > 0$,
the {\rm RS} of \pref{ulineq} is nonnegative if
\[
  \gs \ge \gs_{\min} = \max \, \left[
   \left. \left( \, 
   \sqrt{ \, {\tilde b}^2 -  \,u_g^* \, {\tilde c} } 
   - {\tilde b} \, \right) \right/  u_g^* \, \right] \, .
\]
(2) If $ \,u_j^* = 0$, 
the {\rm RS} of \pref{ulineq} is nonnegative if
$\gs \ge \gs_{\min} = - \max ({\tilde c} / 2 \,  {\tilde b})$.
\vrule height8pt width3pt
\end{lem}

The lemma's limit is very restrictive for large $\dt$, as we now show.
As $\dt \rightarrow \infty$, the terms $a_g$, $\cc$ and $\cc'$ dominate
the definitions of ${\tilde b}$ and ${\tilde c}$.  Hence,
\bey
  \lim_{\dt \rightarrow \infty} 2 \, {\tilde b} 
     & = & a_g \, B_g + \cc' \, u_g^* \; , \nonumber \\
  \lim_{\dt \rightarrow \infty} {\tilde c} 
    & = & a_g \, B_g \, \cc' - a_g \, B_g' \, 
        \cc  \, . \nonumber
\eey
Substituting into the expression for the root and
factoring out $a_g u_g^*$ yields
\[
  \lim_{\dt \rightarrow \infty} \gs_{\min} = \max 
 \frac{a_g}{2} \,
  \left(  
   \sqrt{ ( \ga - \gb )^2 + 4 \gamma } - ( \ga + \gb )
  \right) \, ,
\]
where $\ga = B_g / u_g^*$, 
$\gb = \suml \gklg \Bl' / \rho \cv$,
$\gamma = (B_g'/u_g^*) \, \suml\gklg \Bl' / \rho \cv$
and $\gklg = \gkl / \gk_g$.
The term $a_g = c \, \dt \, \rho \,\gk_g$
equals $\ell_c / \ell_g$,
where $\ell_c$ is the maximum distance a photon can travel in
time $\dt$ and $\ell_g$ is the absorption mean free path for 
the $g$th group.  We now show the remaining expression is of order one.
If the radiation field is at equilibrium,
$\ga = 1$.  The term $\Bl'$ is of order
$\Bl /T$.  If it is exactly
equal to $\Bl/T$, the expression multiplying
$a_g/2$ vanishes.  

If $u_g^* = u_g^0 = 0$ and $p|_{\gs = 1} < 0$, then 
for large $\dt$,
$\gs_{\min} \rightarrow (B_g' \cc - B_g \cc')/B_g$,
which equals $c \dt$ times a term of order one.

We now consider the SI scheme.
As noted above, SI adds the expression $a_g \, B_g' \, (T^* - T^0)$ to the
definition of $2 \, {\tilde b}$.  Effectively, the extra
term means that rather than having $2 \, {\tilde b}$ depend
on the emission source $B_g$ (which is evaluated at $T^*$),
the coefficient depends on the linearization
$B_g + B_g' \, (T^* - T^0)$, with $B_g$ and $B_g'$ evaluated at $T^0$.
If the temperature is decreasing the expression may be negative.
As a consequence, we are not assured that $dp/d \gs$ is positive.
If $u_g^*$ is nonzero, we can find a suitable $\gs$.
However, if $u_g^* = 0$, $p(\gs)$ is a linear function
with possibly a negative derivative.
If that case arises as we query the cells, we set
$\gs = 1$ for the cell in question.
Because of these uncertainties, by default, we run with the FI scheme.

\subsection{Diagonal dominance}
\label{diag}

To prove diagonal dominance, we
compute row sums.  The diffusion terms sum to zero,
since the matrix composed of just these terms
must annihilate the vector $(1, \, 1, \, \ldots)$.\footnote{In extreme cases,
because of finite precision, the diffusion terms may swamp the other terms.
We discuss the possibility in Appendix~\ref{apb}.}  Thus, for
diagonal dominance, 
\[
  \gs + a_g - f_g \sml \al > d \ge 0 \; .
\]
Recalling the definition of $f_g$, the relation is equivalent
to
\[
  0 \le q(\gs) =  \gs^2 + 2{\tilde b} \, \gs + 
    {\tilde c} \; ,  
\]
where
\bey
  2 \, {\tilde b} & = &  a_g + \cc' - d \, , \nonumber \\
  {\tilde c} & = & a_g   \cc' - C_g' \sma 
     - \cc' \, d   \, , \nonumber
\eey
and $\cc'$, $C_g'$ are defined in \pref{cCdef}.
As before, $\gs \ge 1$ is the domain of interest.
The quadratic $q(\gs)$ is
nonnegative for sufficiently large $\gs$.  However,
\[
  q|_{\gs = 1} =   ( 1 + \cc') \, ( 1 + a_g - d \,) -
      C_g' \, \sma \, .
\]
The $a_g$ and $\cc'$ terms are proportional
to $\dt$.  Hence, as $\dt \rightarrow \infty$,
the sign of the expression is dominated by 
sign($\cc' - C_g' \sma$).  Since the expression varies
as $\sml \al ( \Bl' - B_g')$, the sign is indeterminate.
However, $(d\, q / d \gs)|_{\gs = 1}$ 
is positive for $d < 2$.  We have proved:
\begin{lem}
\label{lem:L3}
If $q|_{\gs = 1} \ge 0$ and $d > 0$,
$\ca$ is strictly diagonally dominant for all $\gs \ge 1$.
Otherwise, $\ca$ is strictly diagonally dominant if
\[
  \gs \ge \gs_{\min} = 
    \sqrt{ \, {\tilde b}^2 -  {\tilde c} } 
   - {\tilde b}  \; . \; \vrule height8pt width3pt
\]
\end{lem}

\paragraph{Remark} For large $\dt$, 
\[
  \lim_{\dt \rightarrow \infty} \gs_{\min} = \max 
 \frac{a_g}{2} \,
  \left(  
   \sqrt{ ( 1 - \gb )^2 + 4 \delta } - ( 1 + \gb )
  \right) \, ,
\]
where
$\delta = (B_g' / \rho \cv) \suml \gklg$
and, as before, $\gb = \suml \gklg \Bl' / \rho \cv$,
and $\gk_{\ell,g} = \gk_{\ell} / \gk_{g}$.
As in Lemma~\ref{lem:L1}, when $\dt$ is large,
$\gs_{\min} = \ell_c / \ell_g$ times a term
which should be of order one.

\subsection{Two-step iterative scheme}
\label{errorit}

We've shown that for sufficiently large $\gs$, $\ca$ is an
M-matrix.    Hence,
$(\gll - M_1) - M_2$ and $(\gll - M_2) - M_1$
are regular splittings, and each half of the two-step scheme
\pref{iterh}--\pref{itero} is a convergent iteration
\cite{Varga}, Thm.~3.13, p.~89.
Here we analyze how the scheme reduces the error.
Of particular interest is that for large $\dt$, the scheme
\pref{iterh}--\pref{itero} may {\em not\/} converge 
unless the $\ptc$ parameter $\gs$ is sufficiently
large.

It is convenient to change variables,
\[
  v_j \doteq a_j \, u_j \; .
\] 
The system of interest is then $\ca' v = w$, 
where
\[
  \ca' = \gll - M_1 - M_2
\]
and  $\gll$ is diagonal,
\bey
  \gll_g & = & (\cd_{d,g}/a_g) - f_g + 1 + \gs / a_g   \nonumber\\
  (M_1 \, v)_g & = & \cd_{o,g} \, v_g  / a_g \nonumber\\
  (M_2 \, v)_g & = &  f_g \smlp  \vl \; . \nonumber 
\eey  
If $\ei = v - v^{(i)}$ defines the error for
\pref{iterh}--\pref{itero}, the error satisfies
\bey
  (\, \gll - M_1 \, ) \, \eih & = & M_2 \, \ei \nonumber \\ 
  (\, \gll - M_2 \, ) \, \eio & = & M_1 \, \eih \label{itero2} \; .
\eey
We express the error as a product of 
spatial and frequency components.  For 
a 2D spatial domain,
\be
  \ei_{k,m,g} =  \gepsi_g \, 
    e^{\sqrt{-1} \, ( k \theta_k + m \theta_m) } \; ,   \label{epskj}
\ee
where the indices $k$ and $m$ refer to distinct spatial 
axes.  We now analyze the iteration error
\[
  \eio = ( \gll - M_2 )^{-1} \, M_1 \,
  ( \gll - M_1 )^{-1} \, M_2 \, \ei \; . 
\]

Consider the unit vector ${\hat {\bf e}}_{\ell}$ consisting of
$N$ components, with unity in the $\ell$th position and zeros
for the rest.  Since the initial error $\ei$
is a linear combination of such vectors, it suffices to
analyze the case when the frequency component of 
$\ei$ equals ${\hat {\bf e}}_{\ell}$.  We will prove that
for a properly chosen $\gs \ge 1$,
\[
  || \eio ||_1 \le \zeta  < 1 \; .
\]
In other words, if $\gs$ is sufficiently large,
one iteration of the two-step scheme reduces the error,
which we will show occurs for
$\zeta' < \zeta$, $\gs(\zeta') > \gs(\zeta).$
However, larger $\gs$ denote a smaller $\ptc$
time step, resulting in a longer pseudo-time to reach the
desired steady-state.  

Assuming that the diffusion
coefficient does not vary in space\footnote{Since, as we show,
the worst error arises for spatially constant error, we are free
to ignore the diffusion flux limiter in the analysis.}
and that we use a uniform
2D spatial mesh with mesh size $h$, the error after the first
half step is
\be
  \gepsih_g = f_g \, / \, 
  [ \, 1 - f_g + (\gs / a_g) + 2 \eta_g \,
    ( 2 - \cos \, \theta_k - \cos \, \theta_m ) \, ] \, ,
      \label{epsheq} 
\ee
where
\[
  \eta_g = D_g / a_g h^2 = l_{g,D} \, l_{g,a} 
  / 3 h^2 \, , 
\]
$l_{g,a} = 1/ \rho \gk_g$ is the absorption mean free path,
and $l_{g,D}$ is the diffusion mean free path; the latter is
the sum of the absorption and scattering opacities. 
In \pref{epsheq}, the expression multiplying $\eta_g$ is
nonnegative.\footnote{The corresponding expression in
1 and 3 spatial dimensions is also nonnegative and bounded
by 1.0 and 3.0, respectively.}
Since $f_g < 1$, $\gepsih_g$ is nonnegative.
Assuming the worst case $\theta = 0$ yields,
\be
  0 <  \gepsih_g \le f_g \, / \, ( \, 1 - f_g + \gs / a_g \, ) \; ,
  \label{gepsihlim}
\ee
a result which also holds in 1 and 3 dimensions.
The bound is {\em sharp}; i.e., $\gepsih$
equals the bound if the original error $\ei$
has no spatially varying component.

Since \pref{epskj} holds for $i$, $i+1/2$, and $i+1$, we
now analyze the second half step.
In 2 dimensions,
\[
  M_1 \, \eih = ( \cos \, \theta_k + \cos \, \theta_m ) \,
   2 \, \eta_g  \, \eih \, .
\]
In $n = 1$, 2 or 3 dimensions, the parenthetical expression
contains 1, 2 or 3 cosine terms.  If we again assume $\theta = 0$,
the expression is bounded by $n$.

To determine $\eio$ from \pref{itero2} we invert
$( \gll - M_2  )$ using the Sherman-Morrison formula
by noting that
\[
  \gll - M_2 = \gll' - {\bf f} \, {\bf e}^{{\rm T}} \; ,
\]
where ${\bf e}$ is the vector consisting of all ones,
the components of ${\bf f}$ are the previously
defined $f_g$, and $\gll'$ is diagonal with
\be
  \gll_g' = \eta_g' + 1 + \gs / a_g \; , 
  \;\;\; \eta_g' \doteq 2 \, n \, \eta_g \; .\label{glpdef}
\ee
In \pref{glpdef}, we generalized by allowing for
$n = 1$, 2, or 3 spatial dimensions.  After some algebra,
we obtain
\[
  | \gepsip_g | \le \frac{1}{\gll_g'} \,  
  \left[ \,
     \eta_g' \, \gepsih_g  + 
      \left( \frac{f_g}{1 - {\bf e}^{{\rm T}} \, ( \gll' )^{-1} \, {\bf f} }
      \right) \,
      \sml \frac{\eta_{\ell}' \, \gepsih_{\ell}}{\gll_{\ell}'} \,
  \right] \; ,
\]
where
\be
  1 - {\bf e}^{{\rm T}} \, ( \gll' )^{-1} \, {\bf f} 
          = 1 - \sml \fel / \gll_{\ell}'  \; . \label{elfeq}
\ee
Summing yields the 1-norm,
\[
  || \gepsip ||_1 \le 
  \left[ \, 1 - {\bf e}^{{\rm T}} \, ( \gll' )^{-1} \, {\bf f} \,
  \right]^{-1} \, 
      \sml \frac{\eta_{\ell}' \, \gepsih_{\ell}}{\gll_{\ell}'} \, .
\]
Our task is done if we can show that the RS is bounded by $\zeta.$
Using \pref{elfeq} this entails showing that
\[
  \sml \frac{\eta_{\ell}' \, \gepsih_{\ell}}{\gll_{\ell}'}
    \le \zeta 
    \left( \, 
      1 - \sml \frac{\fel}{\glll'} \, 
    \right) \; .  
\]
After substituting the bound \pref{gepsihlim} and 
simplifying, the inequality becomes
\be
   \smg \left( \frac{C_g'}{\gll_g'} \right)
    \left( \frac{\eta_g'} { 1  - f_g+ \gs/a_g}
    \right)
  \le \zeta 
   \left[  \, \gs +
     \smg \left( \frac{ C_g'}{\gll_g'} \right)
      \, (\gll_g' - 1) \,
   \right] \, . \label{omineq2}
\ee
To summarize, if
\pref{omineq2} is satisfied the two-step scheme
\pref{iterh}--\pref{itero} converges and each iteration
reduces the error by a factor $\zeta$.

We now show that if $\ptc$ is not used, i.e., if $\gs = 1$,
and $\dt$ is large, the scheme may not converge.
If $\gs = 1$, since $a_g \propto \dt$,
$\lim_{\dt \rightarrow \infty} \gll_g' = \eta_g' + 1$
and $\lim_{\dt \rightarrow \infty} f_g = p_g$,
where $p_g > 0$ and $\sum_g p_g = 1$.
Also, if $\gs = 1$, since $C_g' \propto \dt$,
for large $\dt$, the
lone $\gs$ on the RS of \pref{omineq2} is swamped by the sum.
Dividing both sides of \pref{omineq2} by $c \, \dt$, the LS
becomes
\[
  \smg \left( 
     \frac{\rho \, \gk_g \, B_G' \, \eta_g'}{\rho \cv(1 + \eta_g')}
   \right) \, ( 1 - p_g)^{-1} \; .
\]
On the other hand, if $\zeta = 1$, the RS tends to the same sum,
but without the term $(1-p_g)^{-1}$.
This makes the LS larger than the RS,
giving the desired contradiction.  We have proved:
 
\begin{lem}
\label{lem:L5}
If $\gs = 1$ and $\dt$ is large,
\pref{iterh}--\pref{itero} may not converge. \vrule height8pt width3pt
\end{lem}

We now estimate how large to make $\gs$ in order to satisfy
\pref{omineq2}. 
The terms $a_g$ and $C_g'$ are proportional to $\dt$;
also, $\gs \ge 1$ and $\gll_g' > 1$.
Hence, \pref{omineq2} holds for small $\dt$.
To obtain a tractable expression, we derive a relation
that stems from a more stringent inequality.
Equation~\pref{omineq2} holds if we derive a $\gs$
that satisfies a relation insensitive to the lone $\gs$ on the
RS and is obtained by requiring that
the individual terms in the sum satisfy the inequality.
This allows canceling the common term $C_g' / \gll_g'$.
Hence, we seek $\gs$ satisfying
\[
  \eta_g' / ( 1  - f_g+ \gs/a_g) 
  \le \zeta \, (\gll_g' - 1 ) \, .
\]
Recalling that $f_g = C_g'/(\gs + \cc')$ and using \pref{glpdef}
leads to
\[
 0 \le  s(\gs) \doteq \gs^3 + \ga_s \, \gs^2 + 
    \gb_s \, \gs + \gamma_s \; , 
\]
where
\bey
  \ga_s & = & a_g \, (1 +  \eta_g') + \cc'   \, , \nonumber \\
  \gb_s & = & a_g \, [ (1 + \eta_g') \, \cc' - C_g'
      + a_g \eta_g'  \, (1- \zeta^{-1})] \, , \nonumber \\
  \gamma_s & = &  a_g^2 \, \eta_g' \,
      [ ( 1- \zeta^{-1}) \, \cc' - C_g' ]  \, .  \nonumber
\eey
As before, $\gs \ge 1$ is the domain of interest.

To simplify the analysis, we assume $\zeta = 1$, i.e., we 
seek a $\gs$ that guarantees marginal convergence.
To this end, we define
\bey
  {\tilde \gb}_s & = &  a_g \, [ \, 
  (1 +  \eta_g') \, \cc' - C_g'\, ] \;, \nonumber \\
  {\tilde \gamma}_s & = & -  a_g^2 \, \eta_g' \,  C_g' \; . \nonumber
\eey
Consider the cubic
\[
  s(\gs) = \gs^3 + \ga_s \, \gs^2 + 
    {\tilde \gb}_s \, \gs + {\tilde \gamma}_s \; .
\]
For $\gs \ge 1$, all derivatives of $s$ are positive.
If
\[
  s(1) = 1 + \ga_s + 
    {\tilde \gb}_s  + {\tilde \gamma}_s \ge 0 \; ,
\]
then the scheme \pref{iterh}--\pref{itero} converges.
However, if $s(1) < 0$, we need a $\gs > 1$ that renders $s \ge 0.$
To avoid computing cubic roots, we approximate $s$
by a quadratic $w(\gs)$, 
\[
  w(\gs) = (3 + \ga_s) \, \gs^2 +
  ({\tilde \gb}_s - 3)\, \gs + {\tilde \gamma}_s + 1 \; ,
\]
and determine the root of $w$. 
The polynomials $w$ and $s$ and their first two derivatives
agree at $\gs = 1$. 
The difference $s(\gs) - w(\gs) = (\gs - 1)^3$, i.e., 
$w(\gs) < s(\gs)$ for $\gs > 1$.  Hence, the positive root
of $w(\gs)$ overestimates the $\gs$ needed for marginal stability.
We have proved:

\begin{lem}
\label{lem:L6}
If $w|_{\gs = 1} \ge 0$, the scheme \pref{iterh}--\pref{itero} converges.
If $w|_{\gs = 1} < 0$, the scheme converges if
\[
  1 + \tau = \gs \ge \gs_{\min} = 
  \frac{
    \sqrt{({\tilde \gb}_s - 3)^2 - 
      4 \, (3 + \ga_s)\,({\tilde \gamma}_s + 1) }
    + 3 - {\tilde \gb}_s}
  {6 + 2\,\ga_s}  \; . \; \vrule height8pt width3pt
\]
\end{lem}

\section{Multigroup AMR scheme}
\label{mgamr}

In this section, we describe our implementation of
AMR for the multigroup diffusion (MGD)
system.  The scheme necessarily adheres to the code's
general architecture.  That is, on each grid level each physics
module (hydrodynamics, radiation) is called in order.
These comprise the {\em level solves}.
If AMR is used, the code refines in both space and time,
as described by Howell and Greenough \cite{HowGre}.
After a refined level is time-advanced to the next
coarse level time, a synchronization is required in order
to maintain conservation.  For a scalar diffusion equation
and only two levels, coarse and fine, the ``sync-solve''
is difficult enough since it reduces to effectively an
unstructured grid solve over the combined coarse and fine
grids.  For MGD, the difficulty is compounded by having to
sync-solve a coupled system of diffusion equations.

We begin by recalling the equations,
\bey
  \partial_t u_g & = & \nabla \cdot D_g \nabla u_g +
    \gk_g \, ( \, B_g - u_g \, ) \, ,
    \;\; g = 1, \, \ldots,  \, G \label{ugeq} \\
  \cv \partial_t T & = & - \sum_{g=1}^G 
     \gk_g \, ( \, B_g - u_g \, ) \, , \label{emeq}
\eey
where $\cv$ is now the heat capacity, while
$D_g$ and $\gk_g$ are the diffusion and 
coupling coefficients.
For ease of exposition, it is convenient to consider the
one-dimensional case.  The level solve module computes
the solution to
\bey
  \ugi - \ugio & = & (\fgip - \fgim)/h_i + \ggi \, 
     [ B_g(T_i) - \ugi ] \, ,\label{uglev} \\
  \cvi ( T_i - T_i^0) & = & - \sum_{g=1}^G  \ggi \, 
     [ B_g(T_i) - \ugi ] \, , \label{telev}
\eey
where $i$ is the cell index, 
$\ggi = \dt \, \gk_{g,i}$, and $\fgip$ is the
fluence on the right edge of the $i$th cell,
\[
  \fgip = \dt \, D_{g,i+1/2} \, (u_{g,i+1} - \ugi) / h_i \, .
\]

For simplicity, assume there are only two levels,
coarse and fine.  Since \pref{ugeq}-\pref{emeq} are
reaction-diffusion equations, advanced with
backward Euler temporal differencing,
the discretization is unconditionally stable.
Hence, in the following, in order to simplify the derivation,
we assume that both levels
are advanced with the {\em same\/}
timestep.  However, in the code we also 
time-cycle. 
If $i = 1, \ldots, N$ define the indices of all 
coarse-level cells, let
$j = 1, \ldots, J$ define the indices of the refined
cells and $i = I, \ldots, N$ define the indices of those 
coarse cells which are not refined.
Coarse cells indexed with $i = 1, \ldots, I-1$ are
defined as the ``covered'' cells.
We first update the
entire coarse level, then the fine level.
Both levels require boundary conditions (BC)\@.
The coarse level uses the user-specified BC\@.
In the following example, the refined domain abuts the left
side boundary and consists of $J$ cells.  Hence,
the fine level uses the same BC on the left edge.
The fine cell indexed with
$j = J\,$ lies in the interior of the domain.
We reuse the Howell and Greenough \cite{HowGre} infrastructure
to provide a Dirichlet condition for the cell.  The datum is obtained
by interpolating coarse grid data.
Let $k_j$ and $h_i$ define the mesh widths of the fine
and coarse cells, respectively.
After multiplying by the mesh widths and
summing over all cells and groups, we obtain
\begin{eqnarray}
\lefteqn{
  \sum_{j=1}^J k_j \, 
     \left[ \cvj ( T_j - T_j^0) + 
            \sum_{g=1}^G (\ugj - \ugjo)
     \right] +
         }   \nonumber \\
  & & \sum_{i=I}^N h_i \, 
     \left[ \cvi ( T_i - T_i^0) +  
            \sum_{g=1}^G  (\ugi - \ugio) 
     \right]   \nonumber \\
  & & \;\;\; = \sum_{g=1}^G \left( \, \fgin - \fgio - \dfcfg 
                     \right) \, ,\label{utelev}
\end{eqnarray}
where the last term is the fluence miss-match of the $g$th group
at the coarse-fine interface,
\[
  \dfcfg = \fgJp - \fgIm \, .
\]

The AMR scheme assumes that
the system is linear.  Hence, the emission is
expressed as
\[
  B_g(T_i) = \bgs + \bgsp (T_i - \tbi) \, ,
\]
where $\bgs$ and its derivative with respect to $T$, i.e.,
$\bgsp$, are evaluated
at a temperature $\tbi$, e.g., $\bgs = B_g (\tbi)$. 
For semi-implicit Euler differencing, $\tbi = T_i^0$;
if fully implicit, $\tbi = T_i$.  Either way, because
$\dfcfg$ need not be zero, \pref{utelev} shows that
energy may not be conserved after the two level advances. 
To restore conservation, we introduce the
system for the corrections
\bey
  \ugi' & = & (\fgip' - \fgim')/h_i + \ggi \, 
     [ \bgsp \, T_i' - \ugi' ] + \dfcfg / h_i \, ,\label{ugsnc} \\
  \cvi  T_i' & = & - \sum_{g=1}^G  \ggi \, 
     [ \bgsp \, T_i'  - \ugi' ] \, , \label{tesnc}
\eey
where $\fgipm'$ denote the implicit fluxes; they are functions
of $u_g'$.

Equation~\pref{ugsnc} holds for
all groups $g = 1, \, \ldots,  \, G$.  In
\pref{ugsnc}--\pref{tesnc}, the mesh index $i$ varies over
the coarse cells 
not marked for refinement ($i = I, \ldots, N$)
as well as the fine cells ($j = 1, \ldots, J$).
Following the methodology of \cite{HowGre}, we put 
the fluence mis-match  $\dfcfg$ 
into the coarse cell(s) abutting
the interface of the coarse and fine domains.  

Summing the level-advance and
correction solutions yields conservation.
If $\ugi^* = \ugi + \ugi'$ and $T_i^* = T_i + T_i'$, 
combining \pref{ugsnc}--\pref{tesnc}
with \pref{uglev}--\pref{telev}, multiplying 
by the mesh widths, and summing over cells and groups,
yields the desired conservation relation,
\begin{eqnarray*}
\lefteqn{
  \sum_{j=1}^J k_j \, 
     \left[ \cvj ( T_j^* - T_j^0) + 
            \sum_{g=1}^G (\ugj^* - \ugjo)
     \right] +
         }   \nonumber \\
  & & \sum_{i=I}^N h_i \, 
     \left[ \cvi ( T_i^* - T_i^0) +  
            \sum_{g=1}^G  (\ugi^* - \ugio) 
     \right]   
  = \sum_{g=1}^G \left( \, \fgin^* - \fgio^*  
                     \right) \, . \nonumber 
\end{eqnarray*}

Equations~\pref{ugsnc}--\pref{tesnc} present a formidable
task as it requires solving
a simultaneous system of equations for $(G+1)\bar{N}$ unknowns,
where $\bar{N}$ denotes the number of refined cells plus
the number of coarse cells not marked for refinement.
The grid is effectively unstructured since it combines coarse
and fine discretizations of the domain.
We attack the problem by applying a variant of
the ``Partial Temperature'' scheme \cite{LunWil},
\cite{ShHaKe}.  In this scheme, groups are
assigned a random order.  As we cycle through the
groups, each group computes a correction $u_g'$
and a partial temperature $T_g$.  
Note the group index $g$ for the temperature.  Although the scheme
decouples the groups from each other, the partial
temperature $T_g$ changes
as we cycle through the groups.  To be precise,
for each group, we solve the system
\bey
  \ugi' & = & (\fgip' - \fgim')/h_i +  \nonumber \\
        &   &  \;\;\; \ggi \, 
     [ \bgsp \, \tgi - \ugi' ] + \dfcfg / h_i \, ,\label{ugsnc2} \\
  \cvi  \, (\tgi - \tgim) & = & -  \ggi \, 
     [ \bgsp \, \tgi  - \ugi' ] \, , \label{tesnc2}
\eey
where, as above, the mesh index $i$ ranges over all refined cells
and all coarse cells not covered by the fine grid.
For the group index $g_1$ that we first 
pick, $\tgim = 0$ on the LS of \pref{tesnc2}.
Solving \pref{ugsnc2}--\pref{tesnc2} for $g = g_1$
yields the first partial temperature $T_{g_1}$.
This temperature replaces $\tgim$ on the LS of \pref{tesnc2}
for the second randomly picked group $g_2$.
After cycling through the groups, the last one, $g_G$, gives
the desired corrected temperature, i.e.,
$T_i' = T_{g_G,i}'$. 

If \pref{tesnc2} is summed over all $g$, the LS telescopes
and we obtain,
\[
  \cvi  \, T_i' = - \sum_{g=1}^G  \ggi \, 
     [ \bgsp \, \tgi  - \ugi' ] \, .
\]
Because we have $\tgi$ on the RS instead of $T_i'$,
this is not exactly
\pref{tesnc}.  However, if $\tgi$ doesn't vary too much
as we cycle through the groups, the result is no worse
than one obtained with the (commonly-used) partial temperature (PT) scheme
since we apply PT to only corrections of the level-solve solution.
Cycling through the groups in random order avoids
biasing the deviation since the coupling in
\pref{ugsnc2}--\pref{tesnc2} may lower $T$
for one group while raising it for another.  In any case,
the combined solution ($u_g^*, \, T^*$) is still
conservative.

Equations~\pref{ugsnc2}--\pref{tesnc2} are solved using a
Schur complement.  Since \pref{tesnc2} does not involve
spatial derivatives, we can easily solve for $\tgi$.
After substituting the result into \pref{ugsnc2}, we obtain
a single scalar equation for $\ugi'$, albeit now, on the
unstructured grid composed of coarse and fine cells, viz.,
\[
  \ugi' = (\fgip' - \fgim')/h_i +  \nonumber \\
         \;\;\; \ggi \, \egi \, 
     [ \bgsp \, \tgim - \ugi' ] + \dfcfg /h_i \, ,
\]
where, $\egi = \cvi \, / \, ( \cvi + \ggi \, \bgsp )$.
After solving for $\ugi'$, equation~\pref{ugsnc2} yields $\tgi$.
The fluence
miss-match $\dfcfg$ acts
as a source to the corrections.  For groups
with long mean free paths (mfp) and weak coupling, $\dfcfg$
diffuses over the mesh.  For groups with short mfp
and strong coupling, $\dfcfg$ is spread locally over the
group energy $\ugi'$ and ``absorbed'' into the matter.

Before closing this section, we note
an inconsistency in the above {\em multilevel\/}
scheme, indeed in any
scheme embedded in a multi-physics code like ours, which
advances several  modules (hydrodynamics,
heat conduction, radiation) using operator
splitting.  With splitting, on each level, the modules
are advanced in order. 
For simulations using
hydrodynamics and radiation diffusion and running with
coarse L0 and fine L1 levels,
the order of operations is as follows.
Level L0 first advances hydrodynamics, then radiation.  Next,
if refining by a factor of two,
L1 advances in the order: hydrodynamics, radiation, hydrodynamics,
radiation.  The multilevel solve advances in the same order:
hydrodynamics, then radiation.  This implies that the
radiation multilevel solve uses coefficients, e.g.,
$\rho$, that are not the same as those used by the
radiation level solve modules.  In principle,
one cannot simply add the correction equations to the
level solve equations and claim that the sum satisfies
a consistent set of equations.
Nonetheless, the solution remains conservative.


\section{Simulations}
\label{rapAMR}

This section presents results using the multigroup scheme.
We consider three problems.
In Section~\ref{linwin}, we present a 
test problem with a known analytic solution.  
We compare numerical results with tabular data,
previously published by Shestakov and Bolstad \cite{ShBo}.
Using Richardson extrapolation, we show that our $\ptc$ scheme,
i.e., what we apply on a level, is
second (first) order correct in space (time).  
When running with AMR, the temporal accuracy is first order.
Accuracy of the spatial order depends on the norm used to
measure convergence.  In the most stringent $\infty$-norm,
the order degrades to first, or worse, as shown at the
end of section~\ref{linwin}.
Section~\ref{PTCrobust} develops a variation of the
Section~\ref{linwin} test problem in order to
demonstrate the benefits brought
by $\ptc$.  We do this by running with and without
$\ptc$.  We make several runs, each for only one
timestep.  Runs are made with successively larger $\dt$.
Because of fully implicit differencing, as $\dt \rightarrow \infty$,
the numerical solution should approach the time-independent,
steady-state.
The problem in Section~\ref{hotball} brings everything
together.  We simulate the explosive expansion of a
metal sphere suspended in air.  The expansion is due to
sourcing a large amount of energy in a short time
into the sphere.  Simulations
are done with the code's full functionality, i.e., we
couple all of the physics modules and also use AMR.

\subsection{Linear MGD test problem}
\label{linwin}

In this section we present results for a MGD problem with
a known solution.  Due to the nonlinearity of the 
equations, there are no test problems with analytic solutions.
Thus, to validate and verify our algorithm, we consider
the linearized multigroup equations developed
by Shestakov and Bolstad (S\&B) \cite{ShBo} and compare with
tabular data.  

The S\&B tables present results for a 64-group
discretization of the linearized, nondimensional, 
multifrequency diffusion equations derived by
Hald and Shestakov (H\&S) \cite{HalShe}.
In the following, we briefly derive the 
nondimensional system, describe the test problem,
explain how to set up
the problem in a radiation-hydrodynamic code,
demonstrate the problem's relevance to typical applications
of multigroup diffusion, compare results with an
improved-accuracy table \cite{Bol} (supplied in the Appendix),
and conclude by proving that our multigroup
scheme's convergence is first order in time and second
order in space.

The nonlinear multifrequency H\&S system is derived
by assuming slab symmetry, constant density, 
an ideal gas EOS, and an opacity characteristic of free-free
transitions.  One advantage of the H\&S system is its
nondimensional form, which enables comparing results from codes 
using different dimensional units.  The equations are
obtained by choosing characteristic values for density $\rho_0$,
temperature $T_0$, and inverse mean free path (mfp)
$\gk = \gk_0 / \nu^3$ with $\gk_0= {\rm const}$
and $\nu$ the frequency variable.
Radiation emission is given by
a Wien distribution\footnote{It is noteworthy that
H\&S's choice of opacity and Wien spectrum for $B$
gives the same emission source $\gk \, B_W$ as would be
obtained by including stimulated emission (SE) effects \cite{ZelRai}
and using the Planck function, since SE multiplies
$\gk$ by the factor $(1 - e^{-h \nu / k T})$.
Also note that without SE, the resulting Planck-averaged
gray opacity does not exist; the integral diverges.},  i.e., 
$B_W = B_0 \, \nu^3  \exp( - h \nu / k T)$,
where $B_0 \doteq 8 \pi \, h / c^3$ is the same constant defining the
Planck function.
The inverse mfp appears in both the diffusion,
$D = c/3 \gk$, and the radiation-matter coupling terms, $c \, \gk$.
(The diffusion is {\em not\/} flux-limited.)
The normalization proceeds as follows.
The values $\rho_0$, $\gk_0$, and $T_0$
define the other normalization constants,
\begin{center}
\begin{tabular}{ccc} 
   $\nu_0 \doteq k T_0 / h\, , \;\;$ & $\lo  \doteq \nu_0^3 / \gk_0\, , \;\;$ & 
   $x_0 \doteq \lo / \sqrt{3}\, ,$ \\ 
   $t_0 \doteq \lo / c    \, , \;\;$ & $ u_0 \doteq B_0  \nu_0^3  \, , \;\;$ & 
   $E_0 \doteq u_0 \, \nu_0\, .  $ \\ 
\end{tabular}
\end{center}
By defining nondimensional variables,
$x' = x / x_0$, $t' = t/t_0$, $u' = u / u_0$, $\nu' = \nu / \nu_0$, etc.,
(and dropping the primes)
we obtain the normalized system,\footnote{If instead of $B_W$,
H\&S had used the Planck function,
the factor $e^{- \nu / T}$ in Eq.~\pref{ueqnd}
would be replaced by $(e^{\nu / T} - 1)^{-1}$.  However, 
H\&S would then be unable to form Eq.~\pref{Teqnd}, since the
integral over all $\nu$ (the total emission) diverges---see
prior footnote.}
\bey
  \partial_t u    & = &  \nabla \cdot \nu^3 \, \nabla u + \,
         ( \nu^3 e^{- \nu / T} - u \, ) \, / \,\nu^3 \, , \label{ueqnd} \\
  R \partial_t T  & = &  - T + \int_0^{\infty}  
             (u / \nu^3) \, d \nu \, , \label{Teqnd}
\eey
where the constant
\[
   R = (h/k) \,(\rho_0  \cv  / u_0) 
\]
and $\cv$ is the specific heat.
Henceforth, unless stated otherwise, we
use nondimensional variables.

The H\&S system yields a precise
definition of the multigroup equations since the
group integrals can be computed exactly,
an impossible task for definite integrals of the
Planck function.
Given a group structure
$\{\nu_g\}_{g = 0}^G$, after integrating over groups,
\bey
  \partial_t u_g & = & \bnj^3 \partial_{xx} u_g + p_g \, T -
     u_g / \bnj^3  \, , \;\;\; g = 1, \, \ldots, \, G \label{weq}\\
  R \,  \partial_t T  & = & - T + \sum_{g =1}^G  \,
     u_g / \bnj^3   \,  \label{Teq}
\eey
where $u_g = \int_g u \, d \nu$ and
$\bnj$ is a group's representative frequency.  S\&B define
$\bnj$ as $\sqrt{\nu_g \nu_{g-1}}$ and $\bno$ as $\nu_1/2$
since the lowest group boundary is zero.
The emission coefficients are
\be
   p_g \doteq \exp(- \nu_{g-1} / T ) - \exp(- \nu_{g} / T ) \, .
   \label{pdef}
\ee
If the group structure is broad enough, $\sum_g p_g = 1$.

Equations~\pref{weq}--\pref{Teq} are {\em nonlinear\/}
because of the product $p_g \, T$.
To derive an analytic solution, S\&B follow the
approach of Su and Olson \cite{suol}, \cite{suol2}, which
requires a {\em linear}\/ system since it uses Fourier and 
Laplace transforms.  S\&B linearize
by defining a {\em fixed\/} temperature
$T_f$ and substituting $T_f$ for $T$ in \pref{pdef}.

Except for one item, it is easy to assemble the S\&B linearized
MGD system in a conventional radiation-hydrodynamic code.
Such codes usually allow an ideal gas EOS
and a desired analytic form for the opacity.  One
chooses arbitrary values for $\rho_0$, $\gk_0$, $T_0$, and
picks a specific heat $\cv$ to set $R$.
In our simulations, $\rho_0 = 1.8212111 \cdot 10^{-5}$ g cm$^{-3}$,
$T_0 = 0.1$ keV, and $\gk_0 = 4.0628337 \cdot 10^{43}$ cm$^{-1}\,$s$^{-3}$.
To comply with S\&B, we chose $\cv$ to obtain $R = 1$.
Our $\rho_0$, $T_0$, and $\gk_0$ choices 
were dictated purely by reasons of convenience.
Since we compare with a nondimensional result, other constants
may be used instead.

The subtle item is how to force a code's
spectral emission rate to equal $p_g(T_f) \, T$.
We accomplish the task as follows.
The $g$th group's emission 
is $a_g \,[ B_g + B_g' \, ( \,T - T^* \,)]$,
where $a_g = \dt \, c \, \rho \, \gk_g$ and
$\gk_g$ is the group-averaged opacity.
The terms $B_g$ and $B_g'$ are integrals over the $g$th group,
at temperature $T^*$, of the Planck
function and its derivative w.r.t.\ $T$. 
The integrals are computed by a FORTRAN subroutine, which takes
$T^*$ as an input variable.  For the test problem, we
use a different subroutine, which when called, first defines 
\[
  B_g' = 
    (\bnj \nu_0)^3 \, (8 \pi k / c^3) [\exp(-y_{g-1}) - \exp(-y_{g})] \, ,
\]
where $y_{g} = h \nu_g \nu_0 / k T_f T_0$.
After computing $B_g'$, the routine sets $B_g = B_g' \, T^*$.
In the $y_g$ definition,
$\nu_g$ and $T_f$ are nondimensional, while 
$\nu_0$ and $T_0$ are the normalization constants.
The $(\bnj \nu_0)^3$ term
cancels the $1/\nu^3$ dependence of the opacity.

For the test, we consider S\&B's problem 1.
The nondimensional domain is
$0 < x < X$, where we set $X = 4$.  The initial condition is
$T = 1 (0)$ for $x < (>)\, 0.5$ and $u = 0$ everywhere.
We use symmetry boundary conditions at $x = 0$
and homogeneous Milne at $x = X$, i.e., 
$u_g + (2 \ell_g / 3 ) \, \partial_x u_g = 0$,
where $\ell_g$ is the
mean free path.  We use the same
group structure as S\&B: 64 groups, starting at zero,
with widths increasing geometrically by the factor 1.1.
We set $\nu_1 = 5 \cdot 10^{-4}$ as the width of
the first group.\footnote{A 
misprint in \cite{ShBo} erroneously has
$\nu_1 = 10^{-4}$.}  The test
simulates an initially hot slab of material encased by cold matter.
Since $u$ is initially zero throughout,
the solution evolves by first coupling in the hot subdomain.
As radiation diffuses out, it couples to cold matter
thereby heating it.  Because of the opacity's $1/\nu^3$
dependence, the group's diffusion and coupling rates
differ.

\begin{figure}
\begin{center}
\includegraphics[width=\fsize]{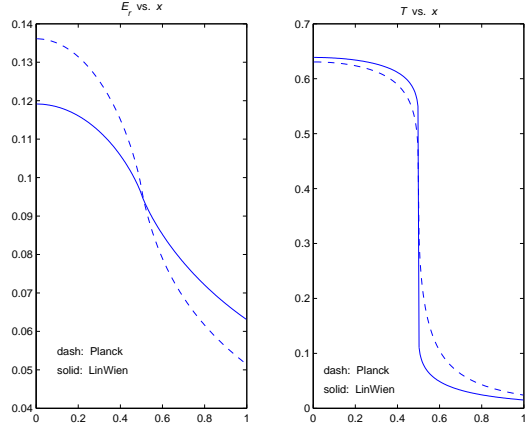} 
\caption{Linear MGD test.  Comparison of the 
linear solution
($T_f = 1.0$) with the solution of the nonlinear MGD system
with Planckian emission; $t = 1.$}
\label{PlLWin}
\end{center}
\end{figure}

Although the problem appears contrived, it
represents effects of radiation diffusion.
We prove the assertion in 
Fig.~\ref{PlLWin}
where we display the temperature $T$ and the
total radiation energy density
$E_r$ $(= \sum_g u_g)$
for two simulations ending at $t = 1$.
Solid lines pertain to the linearized system,
where $T_f = 1.0$.  Dashed lines are solutions of
the ``physical'' nonlinear MGD system using Planckian emission.
The similarity of the solutions validates the relevance of the
test problem.  We used $T_f = 1.0$
(instead of S\&B's $T_f = 0.1$) because over the short 
duration of the simulation, the emission
temperature in the hot subdomain is of order 1.0 rather than 0.1.

We now present our MGD result using S\&B's parameter $T_f = 0.1$.
Table~\ref{relErr} displays the relative errors of $T$ and $E_r$
for various $x$, at $t = 1.0$.  For a variable $f$, we define
the error $\gvep(f) = |(f_x - f_k)/f_x|$, where $f_k$ are our
numerical results and $f_x$ are the S\&B table values,
listed in the Appendix.\footnote{For each point, $f_k$ is the
arithmetic average of the two adjoining cell-centered values.}
\begin{table}
\begin{center}
\begin{tabular}{|ccc|ccc|} \hline
$x$  & $\gvep(T) \cdot 10^3$  & $\gvep(E_r) \cdot 10^3$ & 
$x$  & $\gvep(T) \cdot 10^3$  & $\gvep(E_r) \cdot 10^3$ \\ \hline
0.00 & 0.0016 & 0.3012 & 0.51 & 4.8468 & 0.2785 \\
0.20 & 0.0015 & 0.3028 & 0.52 & 1.8220 & 0.0031 \\
0.40 & 0.0005 & 0.3268 & 0.53 & 1.0528 & 0.1293 \\ 
0.46 & 0.0081 & 0.3903 & 0.54 & 0.7320 & 0.2128 \\ 
0.47 & 0.0174 & 0.4252 & 0.60 & 0.3316 & 0.5263 \\ 
0.48 & 0.0467 & 0.4945 & 0.80 & 0.6099 & 1.3841 \\
0.49 & 0.2205 & 0.6979 & 1.00 & 1.4253 & 2.2138 \\  
0.50 & 0.0019 & 0.3518 &      &        & \\ \hline
\end{tabular}
\caption{Linear MGD test.  Relative errors times 1000.
Numerical result obtained with $T_f = 0.1$, $h = 1/400$,
$\dt = 1/200$. \label{relErr}}
\end{center}
\end{table}
Table~\ref{relErr} shows that we obtain better than 0.5\%
accuracy over the domain $0 \le x \le 1$.  The
worst error 0.48\% occurs for $T$ at $x = 0.51$.
At that point, according to the table in the Appendix,
$T$ undergoes more than a 20-fold drop from its value at
$x = 0.49$.  We focus attention at the domain
near $x = 0.5$ since that is where the variables undergo
the sharpest change.  At these points, we obtain better
than 0.1\% errors, except for $T$ at $x = 0.52$ and 0.53.
Errors near $x = 1.0$ are less important for two reasons.
First, the S\&B domain extends to infinity while ours
extends to only $X = 4$.  Hence, for large $x$, 
our results become less accurate.\footnote{
When we compare results of two simulations at the cells adjoining $x = 1.0$
where one run uses $X = 4$ and for the other, $X = 8$, we find the relative  
differences: $8 \cdot 10^{-6}$, $2 \cdot 10^{-7}$
for $E_r$, $T$, respectively.  Since these
differences are 3-4 orders of magnitude less than the
Table~\ref{relErr} errors at $x = 1.0$,
increasing the domain beyond $X = 4$ would have
little impact on the entries of Table~\ref{relErr}.}
Second, our code requires having a positive min($T$).
Hence, we cannot initialize with $T = 0$ in the cold
region.  At the end of the run, at $x = 1$, our temperature
has risen by only a factor of $10^4$, which
precludes reaching much better than 0.1\% accuracy there.

We were unable to use the S\&B tables for a convergence
study to verify our scheme's convergence properties w.r.t.\
timestep $\dt$ and mesh size $h$.
We speculate that the reason is that 
the truncation
is a mix of errors due to finite $\dt$ and $h$.  Hence, a
refinement study of one may be polluted by an overly
coarse value for the other.  
However, we can use Richardson extrapolation to prove that our
scheme is correct to first order in time and second
order in space. 
Let $v_k$ denote a numerical
solution to an equation discretized by a constant parameter $k$.
For an initial value ODE, $k$ represents the timestep; for a time independent
equation, $k$ is the mesh width.  If $v$ is the analytic
solution,
\[
  v_k = v + \ga \, k^a + \co( k^b) \, ,
\]
where $0 < a < b$, and where $\ga$ is independent of $k$.
In the asymptotic regime, the $k^a$
term dominates the error, which allows ignoring the $\co( k^b)$ term.
Assuming we have three solutions 
$v_k$, $v_{2k}$, $v_{4k}$, a ratio of differences yields
\[
   \frac{v_{2k} - v_{4k}}{v_k - v_{2k}} = 2^a \, . 
\]
The order of convergence $a$ is found by taking logarithms.

\begin{figure}
\begin{center}
\includegraphics[width=\fsize]{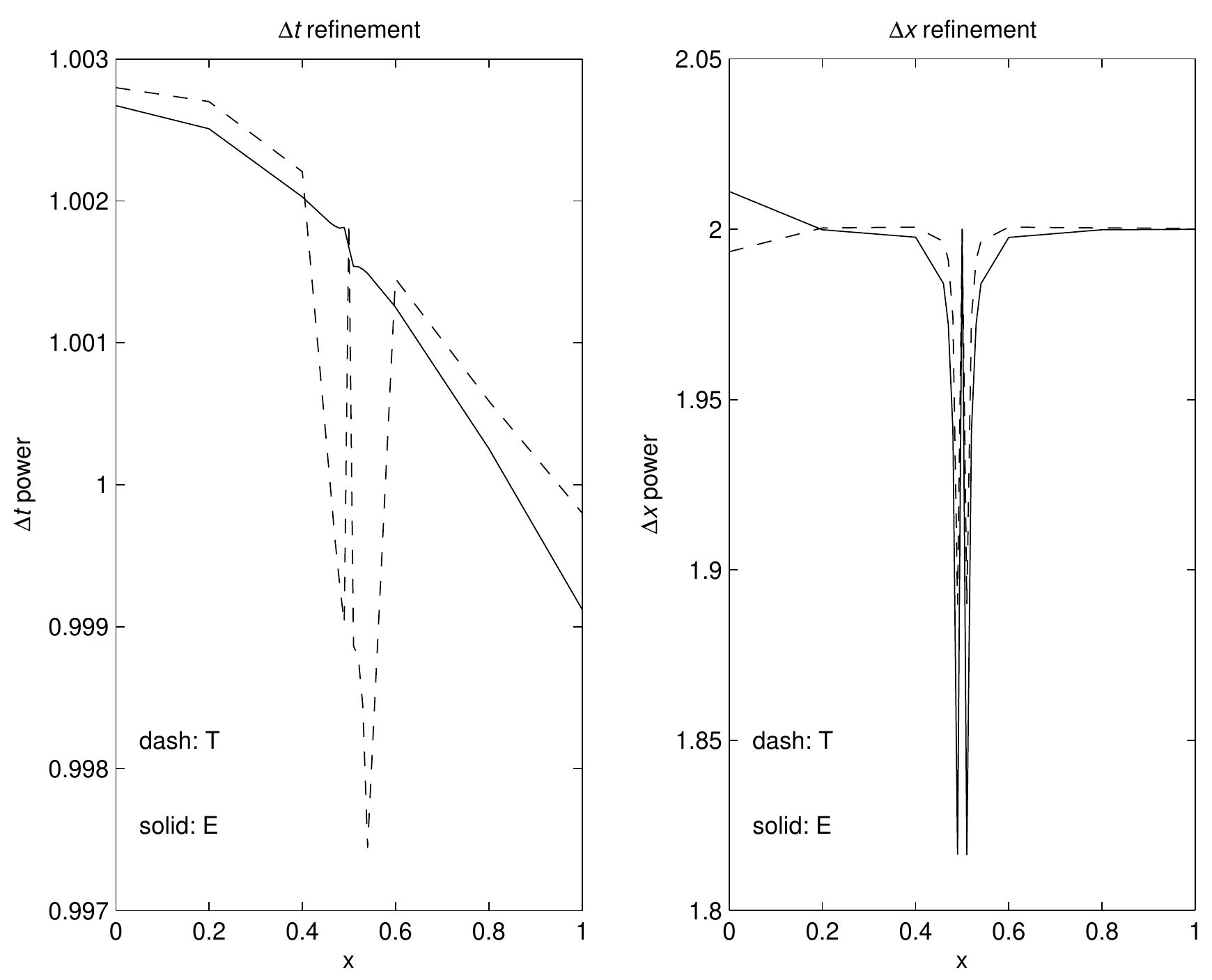}
\caption{Timestep and meshsize orders
of convergence; $\dt$ ($\dx$) on left (right) sides; $t = 1.0$; see text.}
\label{refnmt}
\end{center}
\end{figure}

We apply this procedure to estimate the orders of convergence.
First, for the $\dt$ study, we fix $h = 0.01$ and obtain
three results using $k = 0.5 \cdot 10^{-8}$ s, $2k$ and $4k$.
For the $\dx$ study, we fix
$\dt = 0.5 \cdot 10^{-8}$ s and use $k = 0.0025$.
In both studies, runs are
halted when $t = t_0$.  We compute $a$ at 15 points across
the domain $[\, 0, \, 1]$ for both $E_r$ and $T$ and focus
attention at $x = 0.5$, where the fields undergo
the sharpest change.  Results are presented in
Fig.~\ref{refnmt}.
The left plot clearly displays first order temporal convergence
since $a \approx 1$ across the domain.  The right plot supports
our contention of second order spatial convergence.  The
low $a \approx 1.82$ (1.89) values for $E_r$ ($T$) arise
only at the two points $x = 0.49$, 0.51.  We claim that at
these points, we are not yet in the asymptotic regime.

The results of Fig.~\ref{refnmt} pertain to a solution obtained
on a single level, i.e., without using AMR\@. We now
analyze how AMR affects the order of spatial and temporal convergence.
For each study, $\dx$ and $\dt$, we make three simulations
(as before, we halt at $t = 1.0$)
in order to apply our Richardson extrapolation technique.
In each study, the composite grid consists of a ``base''
level L0 mesh over the entire domain 
and two AMR levels.  Each level refines
by a factor of two.  Both L1 and L2 levels refine around
$x = 0.5$. 
We examine convergence at points $x$ in
{\em all levels.}

\begin{figure}
\begin{center}
\includegraphics[width=\fsize]{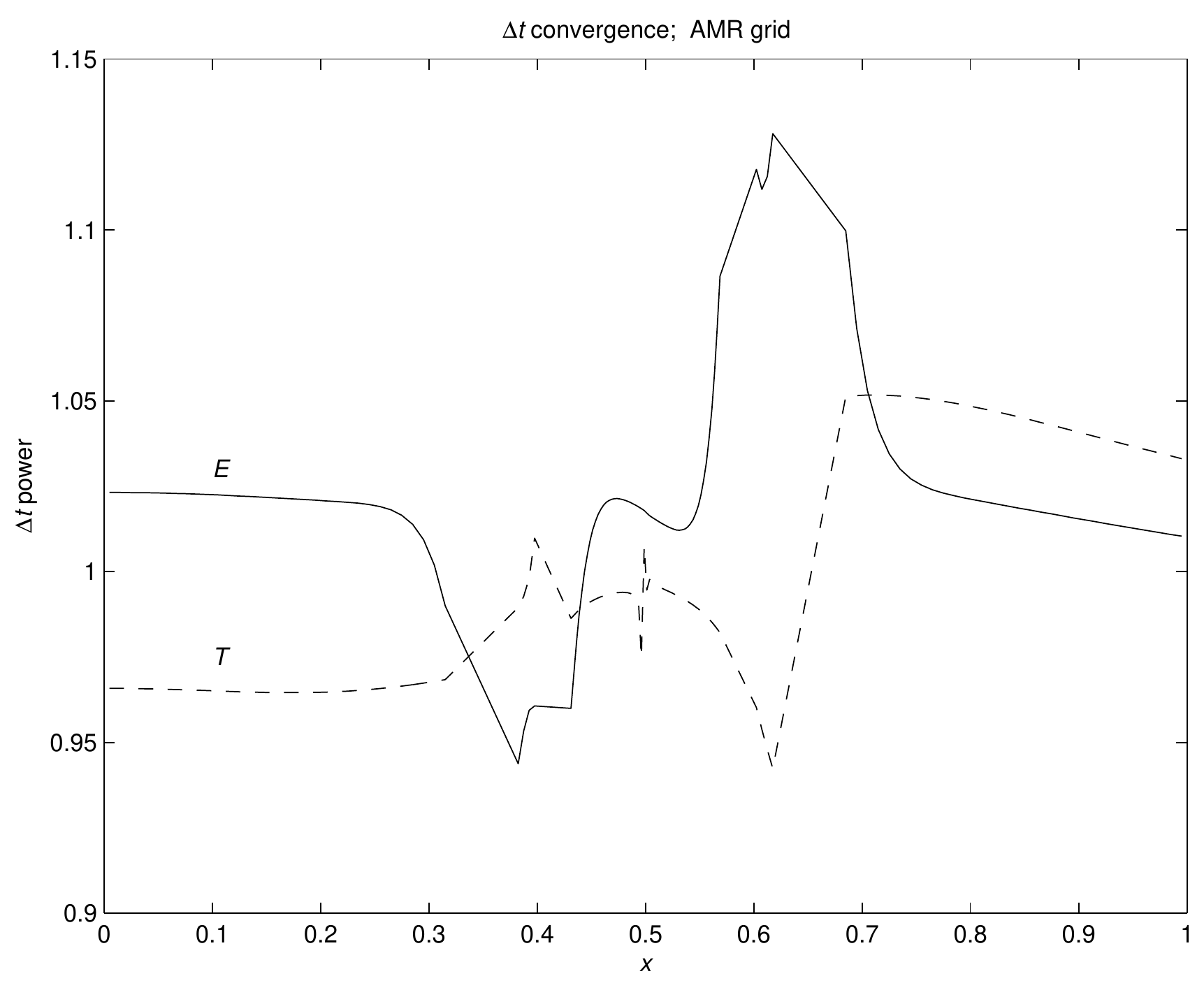}
\caption{Timestep order
of convergence on composite AMR mesh; see text.}
\label{dtrefAMR}
\end{center}
\end{figure}

For the $\dt$ study, all three runs
use the same composite spatial mesh.
We  make three runs;
each with fixed timesteps $\dt_0$, $2\dt_0$ and $4\dt_0$, where
$\dt_0 = 1/400$.  The composite mesh uses
$\dx = 1/100$ on level L0\@.  The L1 mesh extends over
$0.36 \le x \le 0.64$, and the L2 mesh extends over
$0.42 \le x \le 0.58$.  We obtain nearly the same temporal order as for the
level solve.  Figure~\ref{dtrefAMR} displays $a$ for the 128
cells on $0 < x < 1$.   The lowest order,
$a \approx 0.94$, occurs at $x = 0.38$ (0.62) for $E_r$ ($T$)
near the L0 and L1 coarse-fine interface.

\begin{figure}
\begin{center}
\includegraphics[width=\fsize]{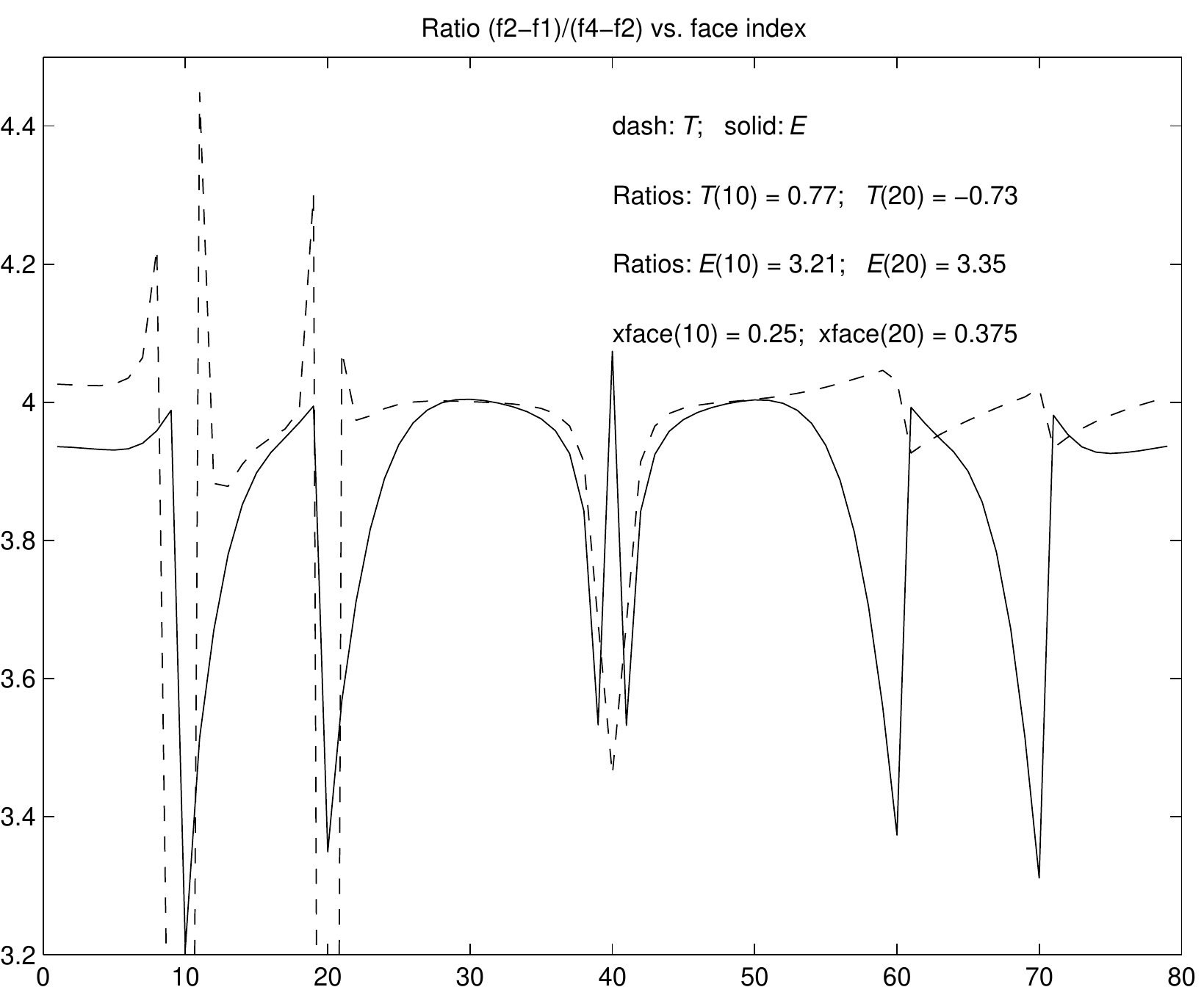}
\caption{Meshsize order
of convergence on composite AMR mesh; see text.}
\label{dxrefAMR}
\end{center}
\end{figure}

The $\dx$ study requires more care.
For each run, the L1 mesh extends over
$0.25 \le x \le 0.75$, and the L2 mesh extends over
$0.375 \le x \le 0.625$.  We refer to the three
runs as R1, R2 and R4, where R1 and R4 use the
``coarsest'' and ``finest'' composite grids, respectively.
For the three runs, the level L0
mesh sizes are 1/40, 1/80 and 1/160, respectively.
Because each AMR level
refines by a factor of two, for R1, the L0, L1 and L2
mesh sizes are also 1/40, 1/80 and 1/160.
The R2 mesh widths are 1/80, 1/160, and 1/320; R4's are
1/160, 1/320 and 1/640.  The composite grids are constructed
so that within each level, the R1 cell boundaries are
also cell boundaries of runs R2 and R4.  
Hence, by arithmetic averaging 
adjoining cell-centered data, we obtain
numerical results at the same points for each run.
These (averaged) values are used for Richardson extrapolation.
Figure~\ref{dxrefAMR} displays the ratio $(f_{R2}-f_{R1})/(f_{R4}-f_{R2})$
for the 79 faces on $0 < x < 1$. 
The ratio is approximately 4 over most of the domain, which indicates
second order convergence.  However at the coarse-fine interfaces, the
order drops significantly; especially for $T$ at $x = 0.25$ and 0.375. 

The loss of accuracy at the coarse-fine (C-F) interfaces is due to
the discretization of the diffusion operator.
We use the infrastructure developed by Howell and Greenough
\cite{HowGre} to assemble the linear systems.  Unfortunately,
the difference stencils---which are not discussed
in detail in \cite{HowGre}---have a shortcoming near the interface.
A more accurate discretization would
yield an asymmetric matrix; for reasons of efficiency,
symmetric linear solvers were preferred.  

The inaccuracy can be analyzed by considering a
derivative such as $u_{xx}$ near the C-F interface.  Assume that level L0 lies to the
left of L1.  For $i = 0$, 1, 2, let $x_i$ denote the first three
cell centers on L1 and let $h$ define the L1 mesh size.
Let $x_c$ denote the center of the coarse cell next to the
C-F interface.  On L1 interior points, e.g., on $x_1$, $u_{xx}$ is 
approximated by the difference:
$(u_0 - 2 \, u_1 + u_2)/h^2$.  Hence, $1/h^2$ is the off-diagonal
matrix coefficient corresponding to $u_0$ on the $x_1$ row.  For the 
matrix to remain symmetric, the $u_1$ coefficient on the $x_0$ row
must equal $1/h^2$.  At $x_0$, $u_{xx}$ is written as
a difference of the right and left fluxes divided by the
cell width $h$.  The right flux is $(u_1 - u_0)/h$.
The left flux is expressed as
the difference $(u_0 - u_c)$ divided by the distance between
the cell centers.  If L1
refines by a factor of two, the distance $x_0 - x_c = 3h/2$.
Thus, at $x_0$, to maintain symmetry, $u_{xx}$ is approximated by
\[ \left.
     \left( \, 
        \frac{u_1 - u_0}{h} - \frac{u_0 - u_c}{3h/2} \,
     \right) 
   \right/ h \, .
\]
Unfortunately, the left flux is not centered on L1's
left-most face (at $x = x_0 - h/2$). 
A Taylor expansion shows that the difference is inconsistent;
it equals $(5/4) u_{xx} + \co(h)$ and this is the source of the error.
However, the error is localized.  In a global sense, it is $\co(h)$,
when computed by integrating over the entire domain:
$\int u_{xx} \, dV$.
This concludes the refinement study on an AMR mesh.

To summarize, in this section we have shown: (a) With a proper choice of $T_f$, 
the test problem mimics MGD physics.
(b) We obtain excellent agreement with the S\&B tables.
(c) Our scheme is correct to first order in time and second
order in space.
(d) On an AMR mesh, the scheme incurs the same loss of accuracy
as the one presented by Howell and Greenough \cite{HowGre} since we use the
same discretization at coarse-fine interfaces.

\subsection{Benefits of \boldmath$\Psi$tc}
\label{PTCrobust}

We now present results that illustrate the benefits obtained
by using $\ptc$.  We show that for sufficiently large
$\dt$, the conventional (ADR) scheme of Axelrod et al \cite{AxDuRh}
i.e., where $\gs = 1$, fails to converge.  Furthermore,
if $\dt$ is only moderately large, so that the ADR scheme does
converge, introducing $\ptc$ accelerates convergence.

We begin by considering a variation of the problem introduced
in Section~\ref{linwin}.  In this section, unless stated otherwise,
we use normalized variables. First, we replace the Wien distribution
with the Planck function.
After normalizing, we obtain an equation similar to
\pref{ueqnd} except that
$e^{- \nu / T}$ is replaced by $(e^{\nu / T} - 1)^{-1}$.
Without stimulated emission effects, the {\em multifrequency\/}
system is ill posed since the RS of the temperature equation
integrates the coupling term over all $\nu$.  (The
integral of $\Bn / \nu^3$ diverges.)  
Since this is only a test, we ignore this complication.
We use seven geometrically spaced groups, whose widths double with increasing frequency.
The leftmost group boundary is zero; the first group width
$\nu_1 = 0.5$; the last boundary $\nu_7 = 63.5$.
As in Section~\ref{linwin}, the first group's opacity
is evaluated at $\nu_1/2$ and the rest are evaluated at the square
root average.
The spatial domain is $0 < x < 2$.
The initial conditions are as before, viz.,
$T = 1$~(0) for $x < (>) \; 0.5$ and $u$ is initially zero.
We impose symmetry boundary
conditions on both left and right endpoints.
Hence, at all times, the total energy should equal
the initial amount $\int_0^{1/2} R \, T \, dx = 1/2$.

\begin{figure}
\begin{center}
\includegraphics[width=\fsize]{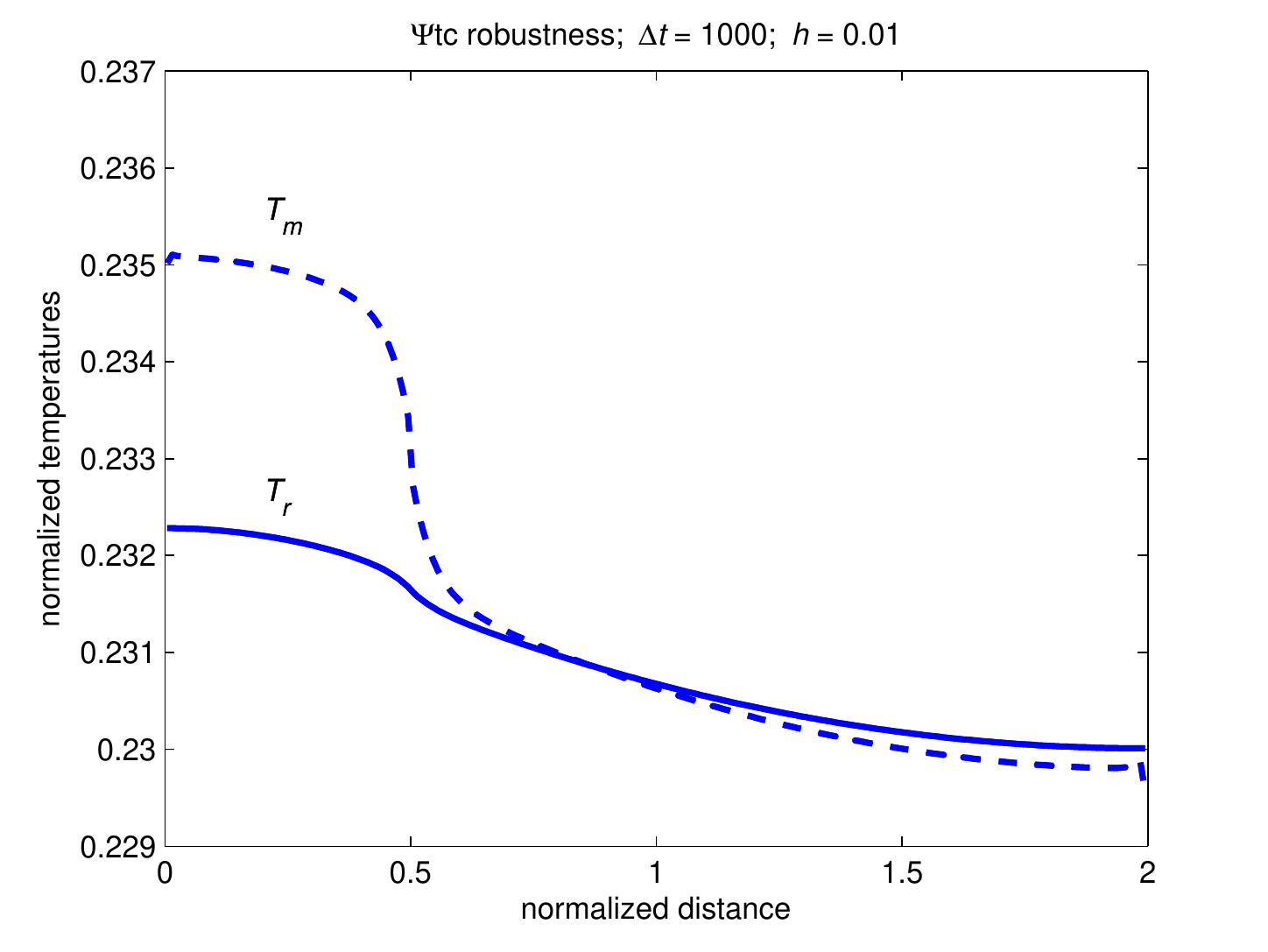}
\caption{$\ptc$ robustness test; solution after one time
advance; $h = 0.01$, $\dt = 1000$.}
\label{PTCrbstEq}
\end{center}
\end{figure}

Our test consists of several runs, each for only one
timestep.  All runs use $h = 0.01$.
We run in fully implicit mode; hence, upon
convergence, the temperature $T$ and emission source $\Bn(T)$
are consistent.  For infinitely large $\dt$, 
a single time advance yields the steady-state with
$T = T_r$, where the radiation energy $E_r = a T_r^4$.
In the nondimensional system, since $\Bn$ is the Planck
function, $a = \pi^4/15$.
Hence, the equilibrium temperature is the solution to,
\[
  2 \, (T_e + a \, T_e^4 ) = 1/2 \, , \;\; 
  {\rm i.e.,} \; \; T_e = 0.2314 \, .
\]
The $\ptc$ result, where $\dt = 1000$,
is displayed in Fig.~\ref{PTCrbstEq}.
The figure shows that the two fields are nearly in
equilibrium and almost spatially constant;
$T_r$ and $T$ vary less than 1\% and 2.4\% respectively. 
The initially high $T$ in $x < 0.5$ has decayed
more than fourfold.  The radiation field, as it
coupled in the initially hot region, diffused
outwards thereby heating the cold region.

The simulations were run with and without $\ptc$.
Both runs consist of nested ``inner'' and ``outer'' loops.
The inner iterations \pref{iterh}--\pref{itero} progress
until the residual and the iterate difference $||\uih - \ui||$
fall below specified tolerances (which may not happen).
At that point, the outer iteration computes a revised temperature $T$
using \pref{dTeq}.  We then
reset $T^* = T$ and use it to recompute the $\Bl$ and $\Bl'$
coefficients.  For the first outer iteration, $T^* = T^0$.
The iterations conclude when the temperature change and the
nonlinear residual fall below their specified tolerances.

The problem's difficulty increases with $\dt$.
Without $\ptc$, it
becomes impossible to solve if $\dt$ is very large because
of the computer's finite precision.
For large $\dt$, the time derivatives, e.g., $(u - u^0)/\dt$,
are dominated by the other terms.  Hence, the initial condition
$( \, u^0, \, T^0)$ becomes less relevant.
Unfortunately, energy conservation depends on 
``remembering'' the initial condition.  
The boundary conditions enhance the difficulty.
If the initial condition is indeed ``forgotten,''
the solution is not unique.  Any equilibrium temperature
$T_e = T = T_r$ is a steady-state.

For runs without $\ptc$, we impose $\gs = 1$ and determine
for which magnitude $\dt$ the iterations fail to converge.
Runs using
$\ptc$ proceed as follows.  We first compute the three different
$\gs$ required to have (1) a nonnegative RS, (2) diagonal
dominance, and (3) convergence of inner iterations.
That is, the $\gs$ must satisfy the lemmas of Sections~\ref{posw},
\ref{diag} and \ref{errorit}.  The iterations commence using
the largest $\gs$.  The parameter $\gs$ is fixed for
each outer iteration. 
Experience has shown that the lemmas
give an overly large $\gs$.  Hence, we use the lemmas to set $\gs$
for only the first outer iteration.  
Subsequent outer iterations decrease $\gs$ as follows.
Recall $\gs = 1 + \tau$ and that only when $\tau = 0$ do we solve
the correct discretization of the equations.
Successive outer iterations multiply $\tau$
by a constant factor, i.e., $\tau \rightarrow \gat \, \tau$.
The factor may be changed by the user.
For small $\gat$, $\tau$ decreases quickly, but the
resulting linear system is harder to solve.
For the hardest test, where $\dt = 1000$ (see below),
we experimented and found better results with $\gat = 0.5$
than with $\gat = 0.25$.

Our tests begin with $\dt = 20$, a magnitude at which both
modes, with and without $\ptc$, converge and give
nearly identical results.  For this moderately large 
$\dt$, $\ptc$ brings the benefit of faster convergence:
37 vs.\ 50 CPU sec, i.e., nearly 33\% faster.
For $\dt = 100$, the two modes still converge and
give very similar results, but they are now at the
limit of convergence.  The $\ptc$ run is significantly faster:
56 vs.\ 205 sec, an almost fourfold improvement.
For $\dt = 200$, the non $\ptc$ run does not converge.
However, its final iterate temperatures still look physical;
$T_r$ is 0.5\% uniformly higher than the
corresponding converged $\ptc$ profile.
Our $\ptc$ implementation has its own limit.
The Fig.~\ref{PTCrbstEq} result, where $\dt = 1000$,
also fails to converge.  Nonetheless, the result is physical
and conserves energy to nearly 11 decimal digits.
Non-convergence is evidenced by small dips
in the matter temperature $T_m$ at the cells abutting the
left and right boundaries.  
At the end points, $T_m$ changes very slowly from one iteration
to the next.  
The iterations effectively stall. 
Although the residuals continue to decrease, they
have such a slow decay that the run halts when it reaches
the iteration limit.
The run without $\ptc$ and $\dt = 1000$ diverges due to
negative internal energies.
To summarize, $\ptc$ not only decreases the runtime
but also brings an extra degree of robustness.

\subsection{Expansion of a hot aluminum sphere} 
\label{hotball}

In our opinion, the hardest aspect of code
development is integrating a module into a
multi-physics code and running ``real'' problems.
For us, this implies simulations of multiple
materials, whose properties are listed in tables,
using hydrodynamics, heat conduction, 
radiation modules, and, naturally, AMR.

For the final test
we consider the following problem.  An Aluminum (Al) sphere
of radius 15.5 cm is suspended in air.
The initial densities are $\rho = 2.68118198$ and 0.00129
g/cm$^3$ for Al and air, respectively.  Both materials are
initially at $T = 375.936$ K.\footnote{Inputs
are tailored so that our EOS returns equal pressures
for both materials, approximately 1 bar.}  There is
initially no radiation energy: $E_r|_{t=0} = 0.$
At $t = 0$, we inject energy into the
radiation field, but only into the domain containing Al.
The energy is added over 0.1 ns, at which point we
have loaded a yield $Y$ (erg) into the problem.
Energy is added with a Planckian spectrum.
Unless stated otherwise, the
simulations presented in this section use two
AMR levels; $h = 2$, 4 cm, and a base grid with $h = 8$~cm.

We compare simulations
in which radiation transport is modeled by a single
diffusion equation for the radiation energy density
(gray diffusion) to runs where the transport is
modeled with multigroup diffusion (MGD).
We describe results where $Y \approx 11$ kT and 
$Y \approx 1$ MT.
\footnote{Using the conversion $4.18 \cdot 10^{19}$ erg/kT,
the actual yields are
10.9731 kT, 0.9870682 MT, 10.9665 kT, and 0.9862604 MT
for the two gray and two MGD runs, respectively.}

The problem simulates a strong explosion in air;
the parameter choice corresponds to a nuclear source.
The effects are well-known:
Zel'dovich and Raizer \cite{ZelRai}
Ch.~IX, Brode \cite{Bro}, Landshoff \cite{Lan}.
Initially, radiation dominates the dynamics:
a fast thermal wave propagates through the surrounding air.
When the wave slows to sonic speeds
(of the hot air), the steep pressure gradient
gives rise to a strong shock.  Finally, hydrodynamics
dominates.  Salient effects are similar to the simulation
of a point explosion using hydrodynamics and nonlinear heat conduction
(Shestakov \cite{ShePtEx},
``Non-Self-Similar-Problem'' section).

Before presenting our results, we summarize them. 
For the lower yield, gray and MGD simulations
are very similar.  However, for $Y = 1$ MT, 
the gray and MGD simulations differ
significantly and this, we feel, is a new result.
Although it contradicts established theory
(Brode \cite{Bro}) we believe it to be correct
since it is explained by examining spectra of the
radiation field (see below).  
Furthermore, our
MGD result is corroborated by the trusted computer code
LASNEX \cite{las}.

\begin{figure}
\begin{center}
\includegraphics[width=\fsize]{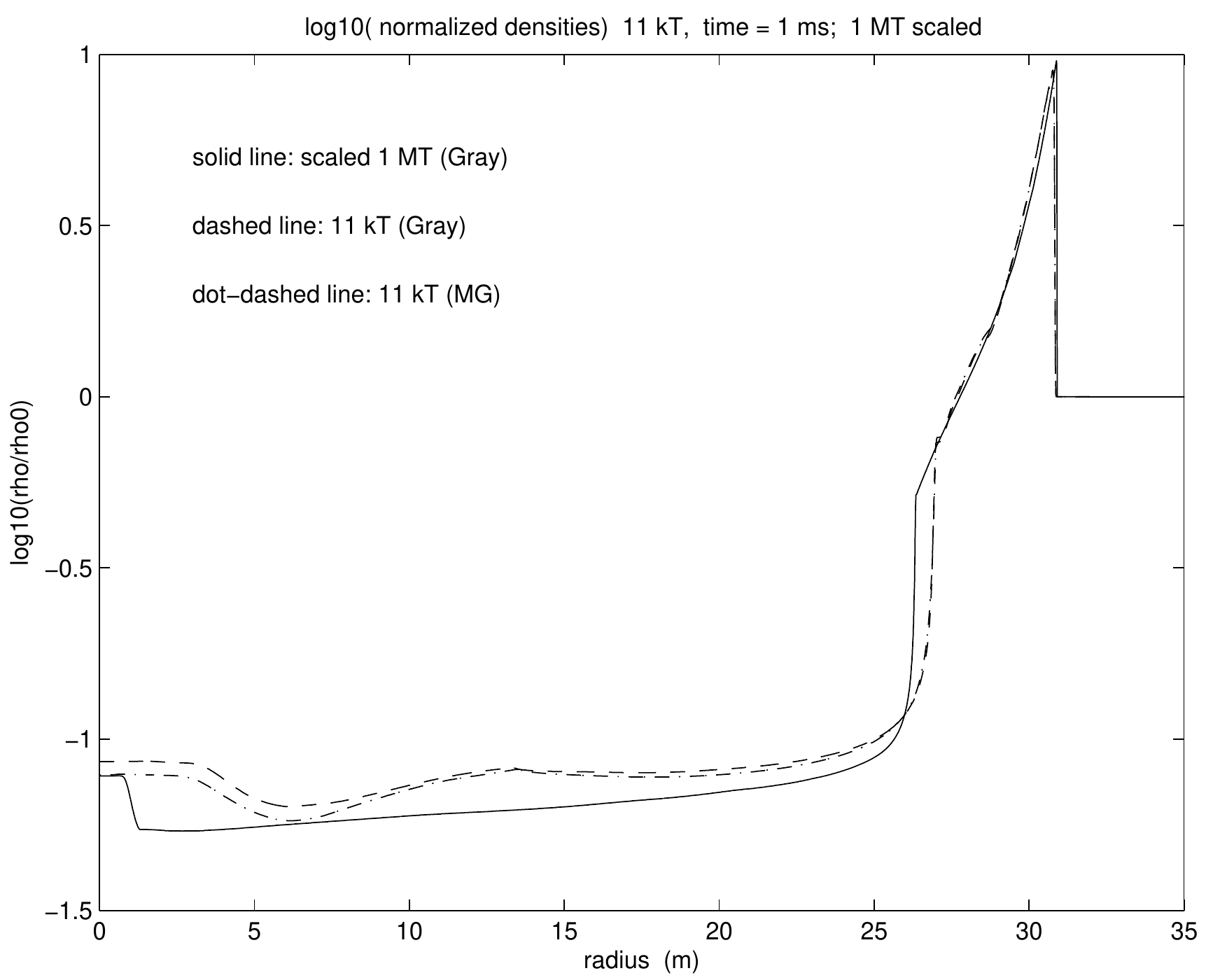}
\caption{Hot sphere problem.
Log of normalized densities; $Y = 11$ kT yield, $t = 1$ ms,
gray and multigroup diffusion;  $Y = 1$ MT gray curve is scaled.}
\label{rho_11kT}
\end{center}
\end{figure}

\begin{figure}
\begin{center}
\includegraphics[width=\fsize]{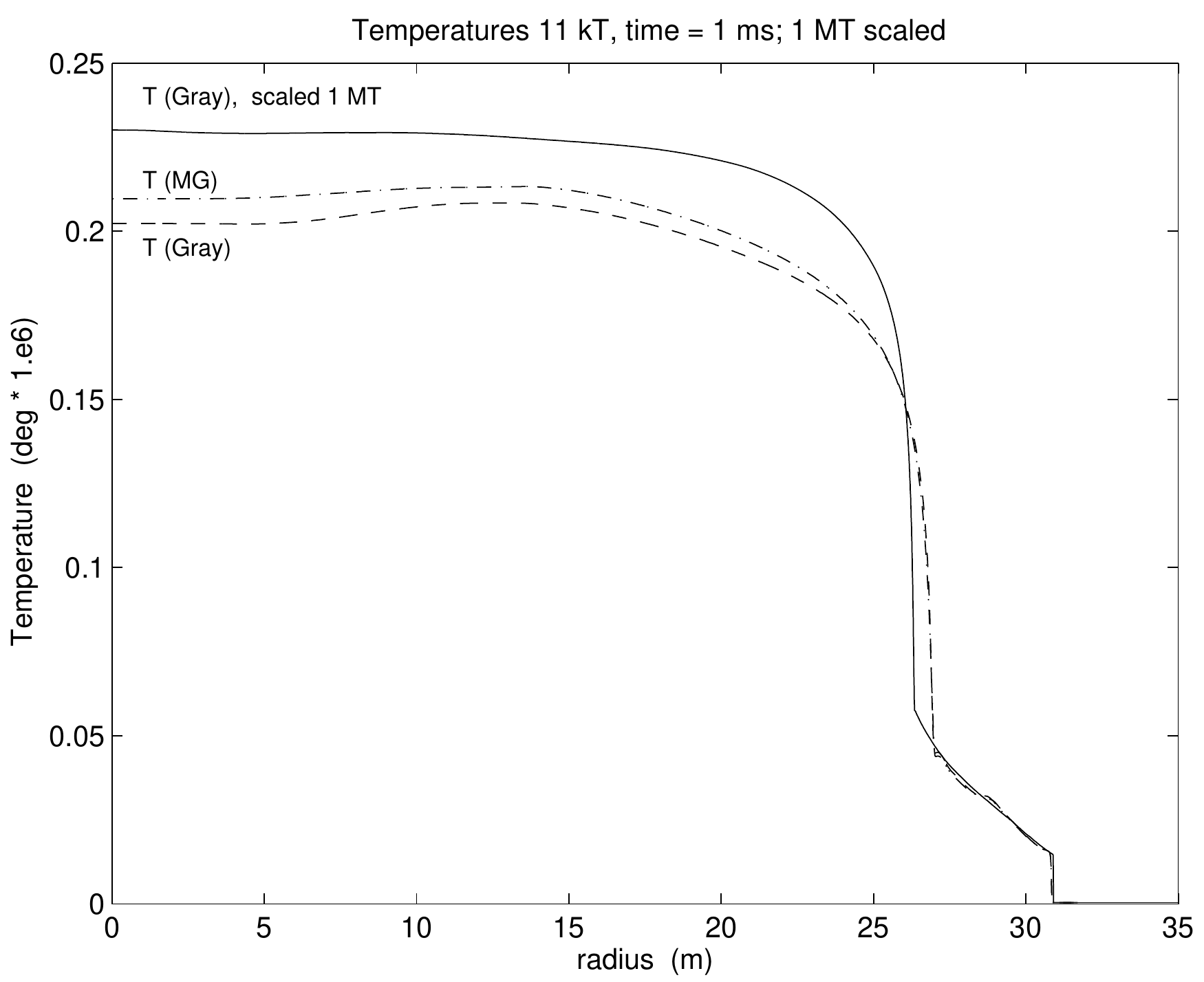}
\caption{Hot sphere problem.
Matter temperatures $T$; $Y = 11$ kT yield, $t = 1$ ms,
gray and multigroup diffusion;  $Y = 1$ MT gray curve is scaled.}
\label{t_11kT}
\end{center}
\end{figure}

\begin{figure}
\begin{center}
\includegraphics[width=\fsize]{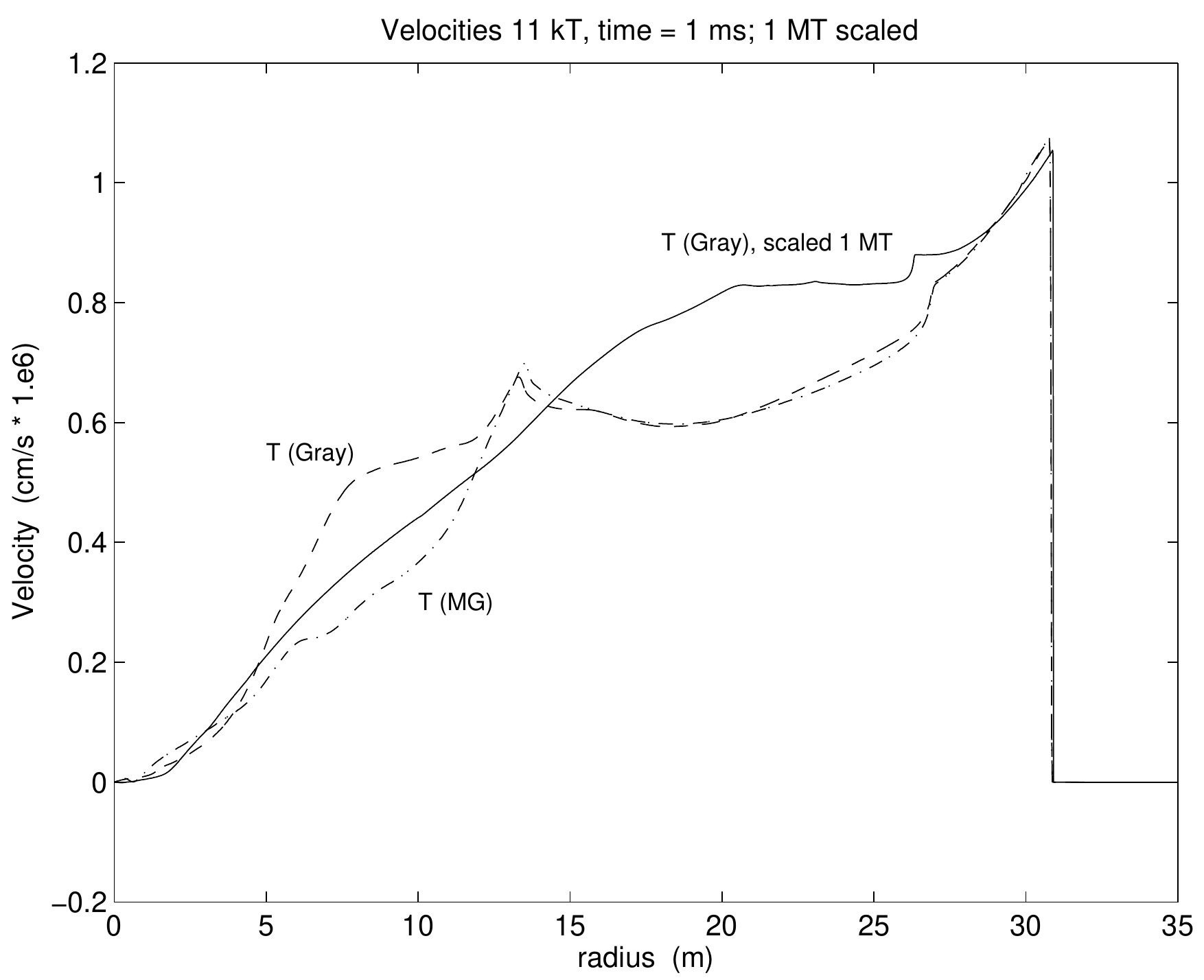}
\caption{Hot sphere problem.
Velocities; $Y = 11$ kT yield, $t = 1$ ms,
gray and multigroup diffusion;  $Y = 1$ MT gray curve is scaled.}
\label{u_11kT}
\end{center}
\end{figure}

Figures~\ref{rho_11kT}, \ref{t_11kT} and \ref{u_11kT}
display densities, temperatures and velocities, respectively.
Each figure contains three curves.
Two are from gray and MGD simulations
with $Y = 11$~kT\@.  The third curve is from
a simulation using gray diffusion and a yield 
$Y = 1$ MT\@.  The 1 MT 
curves are drawn after implementing Sachs scaling, i.e., by 
scaling time and radii by the cube root
of the yield ratio $R_Y = (Y_1/Y_2)^{1/3}$, where
$Y_1 = 11$ and $Y_2 = 1000$.  Hence, while the $Y = 11$
results are taken at $t = 1$ ms, the 1 MT results are at
$t = 4.48$ ms and the 1 MT radii have been
divided by $R_Y$.  Figure~\ref{rho_11kT} displays
$\log_{10} (\rho / \rho_0)$, where $\rho_0 = 0.00129$~g/cm$^3$ is
the ambient air density.  
Although the close agreement displayed in Figs.~\ref{rho_11kT},
\ref{t_11kT} and \ref{u_11kT} may not surprise, it is indeed
remarkable how well the gray scaled 1 MT curves
compare with the lower yield results. 
The similarity of the $Y = 11$ kT
gray and MGD curves indicates
that gray diffusion is adequate for small $Y$.

The results in Figs.~\ref{rho_11kT}, \ref{t_11kT} and \ref{u_11kT}
are characteristic of an event transitioning from a radiation dominated
regime to one dominated by hydrodynamics.
Figures~\ref{rho_11kT} and \ref{t_11kT} depict a strong shock
at $r = 31$~m separating from a fireball of radius 26-27~m.

In order to validate our gray $Y = 1$ MT simulation, we
continue the run to $t = 7$ ms and find good 
qualitative agreement when we compare with Brode \cite{Bro}.  
Quantitatively, at $t = 7$ ms,
we find a strong shock at $r = 164$ m, whereas
Brode finds it at $r \approx 190$ m.  Both simulations show a nearly
tenfold density rise at the shock, while inside the
fireball, $\rho \approx 5 \cdot 10^{-5} $ cm$^3$.  For the central
($r = 0$) temperature
we have $T = 2.04 \cdot 10^5$ K at $t = 7$ ms
vs.\ $\approx 2 \cdot 10^5$ K for Brode.
Our fireball radius is 138~m ($\approx 160$ for Brode),
and our shock temperature is $1.65 \cdot 10^4$ K
($\approx 1.6$--$1.7\cdot 10^4$ K for Brode).

\begin{figure}
\begin{center}
\includegraphics[width=\fsize]{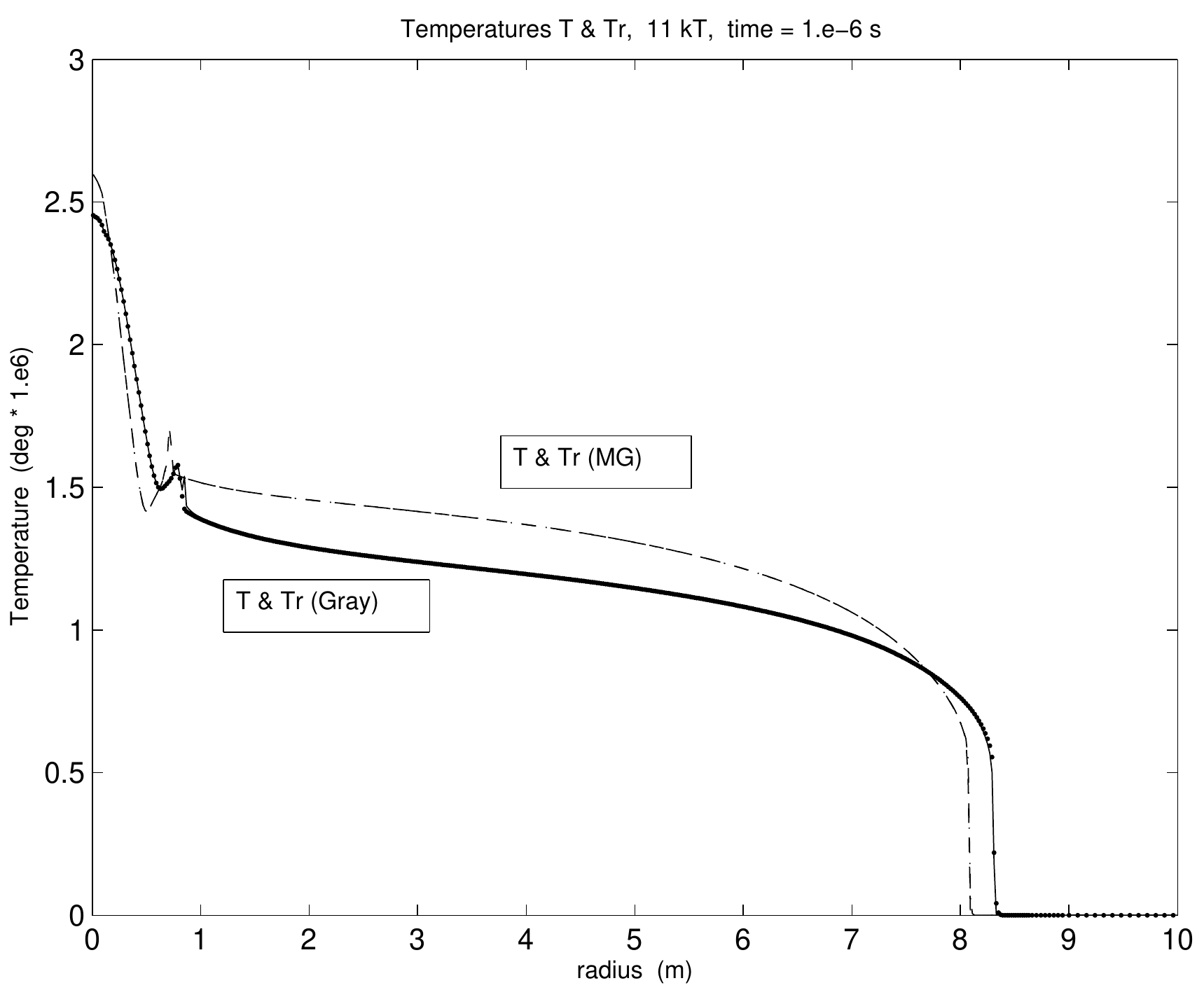}
\caption{Hot sphere problem.
Gray and multigroup temperatures $T$ and $T_r$,
$Y = 11$ kT, $t = 1 \mu$s.}
\label{trt_11kT}
\end{center}
\end{figure}

We now compare the gray and MGD results for $Y = 11$ kT yield
at the earlier time, $t = 1 \; \mu$s,
when the solution is dominated by radiation.
At this time, since the thermal wave is supersonic,
it suffices to only examine the temperatures $T$ and $T_r$,
where, for both gray and MGD simulations,
$T_r \doteq (E_r/a)^{1/4}$ and $a$ is the radiation constant.
(Although the Al ball has ballooned to nearly 1 m, which
launches a strong shock at the Al/air interface,
there is little separation between the interface and the shock.
Thus, beyond 1 m,
the air density is nearly the same as it was initially.)
Figure~\ref{trt_11kT}, which displays the temperatures,
shows little
difference between gray and MGD\@.  Both models
display a fireball extending to $r = 8.1$-8.4~m
and a central $T \approx 2.5 \cdot 10^6$~K; both also 
display the start of the shock at the Al/air interface,
as evidenced by the spike at $r \approx 0.8$ m.   
  
\begin{figure}
\begin{center}
\includegraphics[width=\fsize]{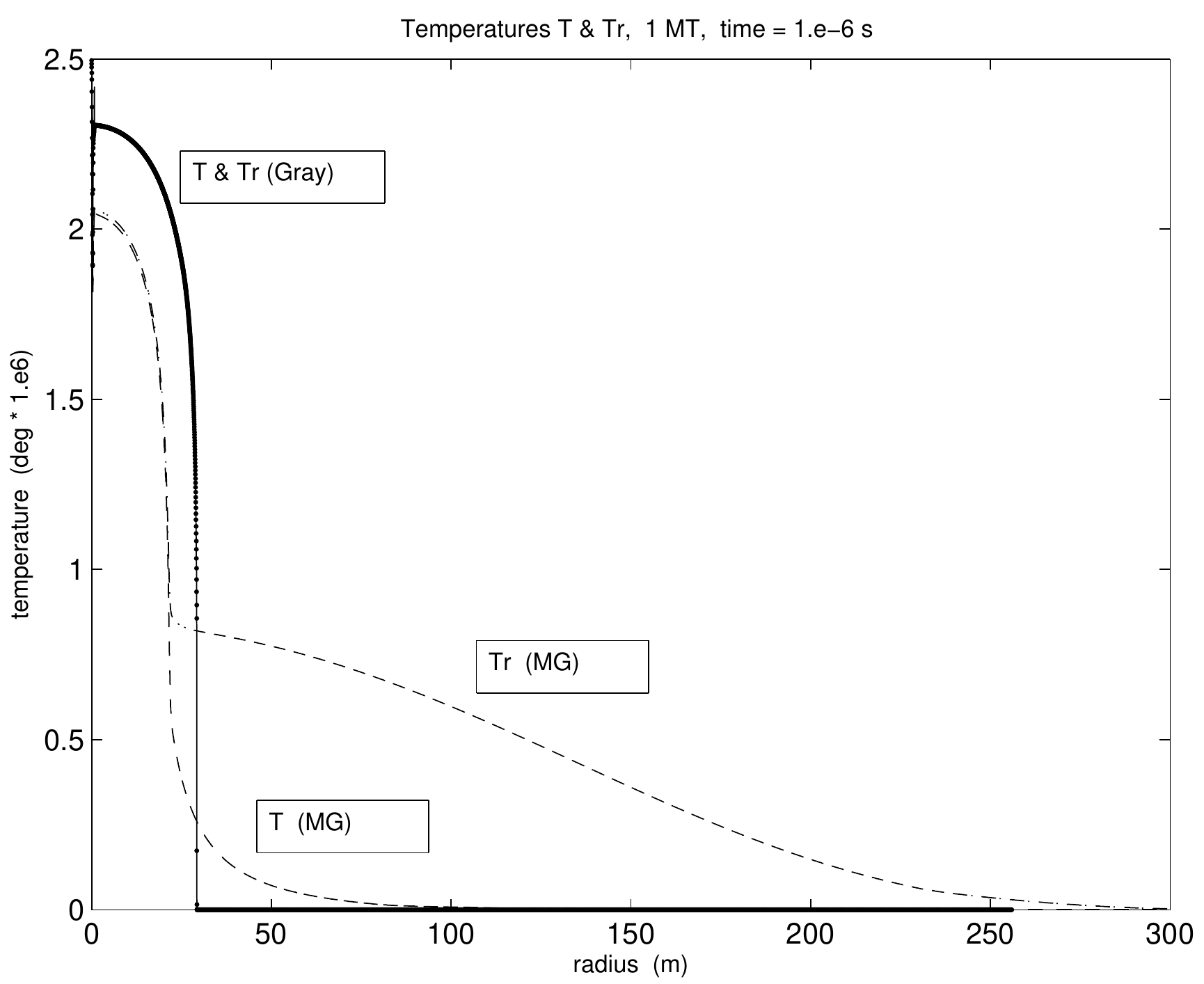}
\caption{Hot sphere problem.
Gray and multigroup temperatures $T$ and $T_r$,
$Y = 1$ MT, $t = 1 \mu$s.}
\label{trt_1MT}
\end{center}
\end{figure}

However, for high yield, the
gray and MGD simulations differ dramatically.
Figure~\ref{trt_1MT} displays $T$ and $T_r$
for $Y = 1$~MT at $t = 1 \mu$s.\footnote{The spatial scale of
Figure~\ref{trt_1MT} cannot resolve the small, but nevertheless
significant hydrodynamic effects which expand the Al sphere to
$r \approx 80$~cm.  For MGD, the temperature is not monotone
w.r.t.\ to $r$ near the origin.  It falls from a central value
of $2.6 \cdot 10^6$~deg to $1.8 \cdot 10^6$ at the edge of the
sphere (due to the rarefying Al) then rises to $2.06 \cdot 10^6$
in the air.}
We see that for gray diffusion,
$T = T_r$; just as for $Y = 11$~kT\@.
The gray diffusion thermal wave,
which is still supersonic,
has a front at $r \approx 30$ m.
However, the MGD result is strikingly different.
Multigroup diffusion lowers the central temperatures
by more than 10\%.  More surprisingly,
for MGD, $T$ and $T_r$
are tightly coupled only out to $r \approx 20$ m.  Beyond that,
at $T \approx 8.5 \cdot 10^5$ K, $T$ and $T_r$ decouple.
The radiation temperature 
extends to $r \approx 300$~m, which is the
free-streaming limit. 

\begin{figure}
\begin{center}
\includegraphics[width=\fsize]{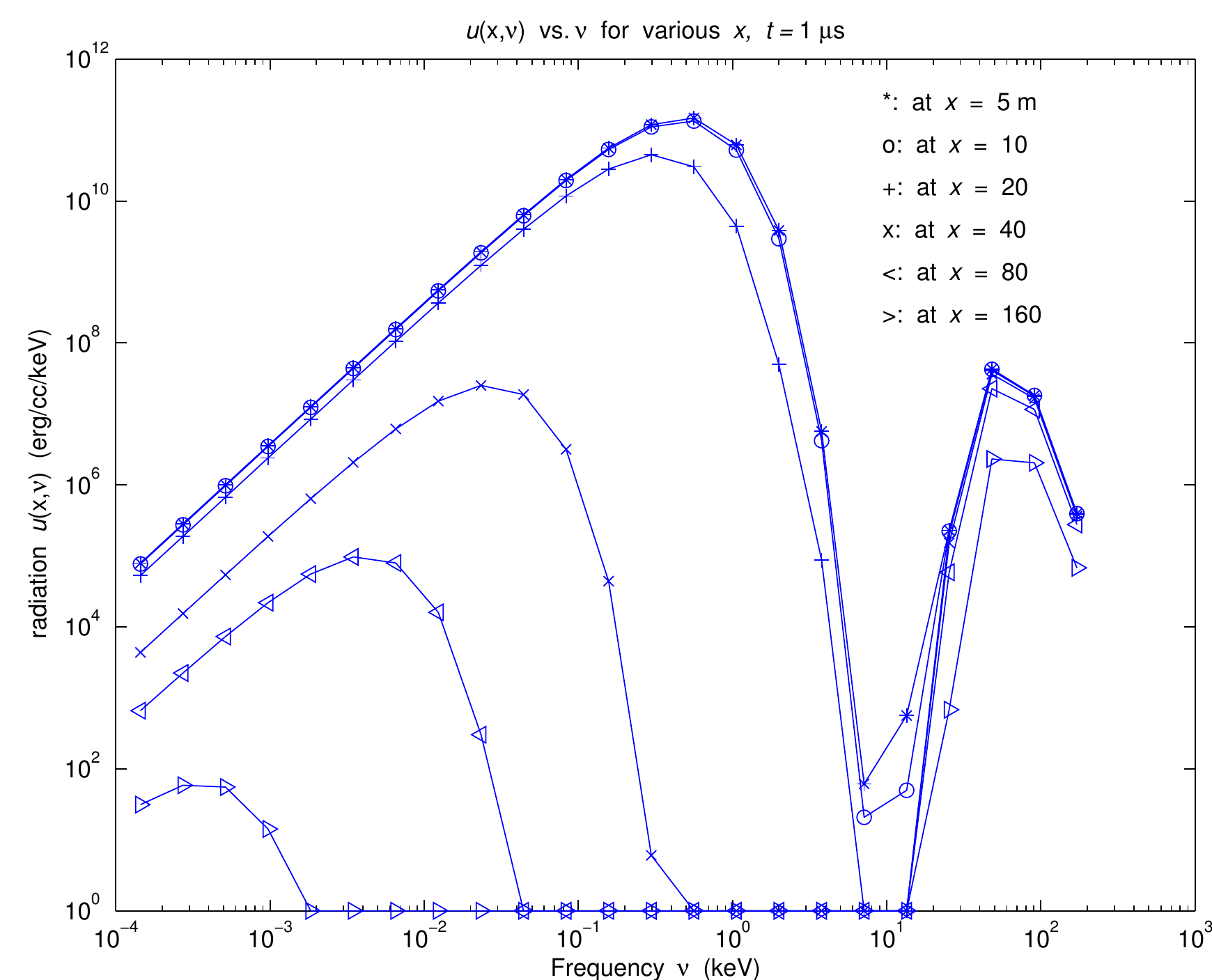}
\caption{Hot sphere problem.
Spectral radiation energy (erg/cc/keV) vs.\ frequency (keV)
at various radii; multigroup physics only;
$Y = 1$ MT, $t = 1 \mu$s.}
\label{unuB}
\end{center}
\end{figure}

To examine why the high yield
gray and MGD simulations differ, we turn off hydrodynamics
and heat conduction, repeat the simulation,
and find temperatures similar to
Fig.~\ref{trt_1MT}.  This is not surprising
since the dynamics are radiation-dominated.
To gain more insight, we examine spectra.
Figure~\ref{unuB} displays
the spectral radiation energy vs.\ frequency 
at 5--160~m.
Evidently, the frequency-dependent air opacity is
responsible.  High frequency (30--200 keV) photons travel
largely unimpeded whereas near the
origin, the spectrum develops a hole at 10 keV\@.
Moving away from the center, the hole
progresses to lower frequencies so that at 100--200~m, the spectrum
consists of two peaks, one at the high frequencies, another
near the visible range.  Since the latter 
contains little energy, the protruding radiation ``tongue''
of Fig.~\ref{trt_1MT} is due to the high frequencies.

We believe that the difference between the $Y = 11$~kT and $Y = 1$~MT 
MGD simulations is due to the factor of 100 between the yields.
Because the energy is added with a Planckian
spectrum, the initial maximum temperatures differ by roughly
the fourth root, or approximately 3.  Since the initial
temperatures are of order 3--5~keV, the high frequencies
have a nearly Wien distribution, $\nu^3 \, e^{-\nu/T}$.
Hence, we expect the $Y = 11$~kT spectrum to be
$e^{-\nu/T}\!/e^{-\nu/3T}$ or $e^{-2 \nu/3T}$ times
smaller than the high yield case.
Substituting $T = 3$ and $\nu = 100$~keV gives a very small
number.  The conclusion is that the $Y = 11$~kT case has
an insignificant number of those energetic photons that
are not absorbed by air.

\begin{figure}
\begin{center}
\includegraphics[width=\fsize]{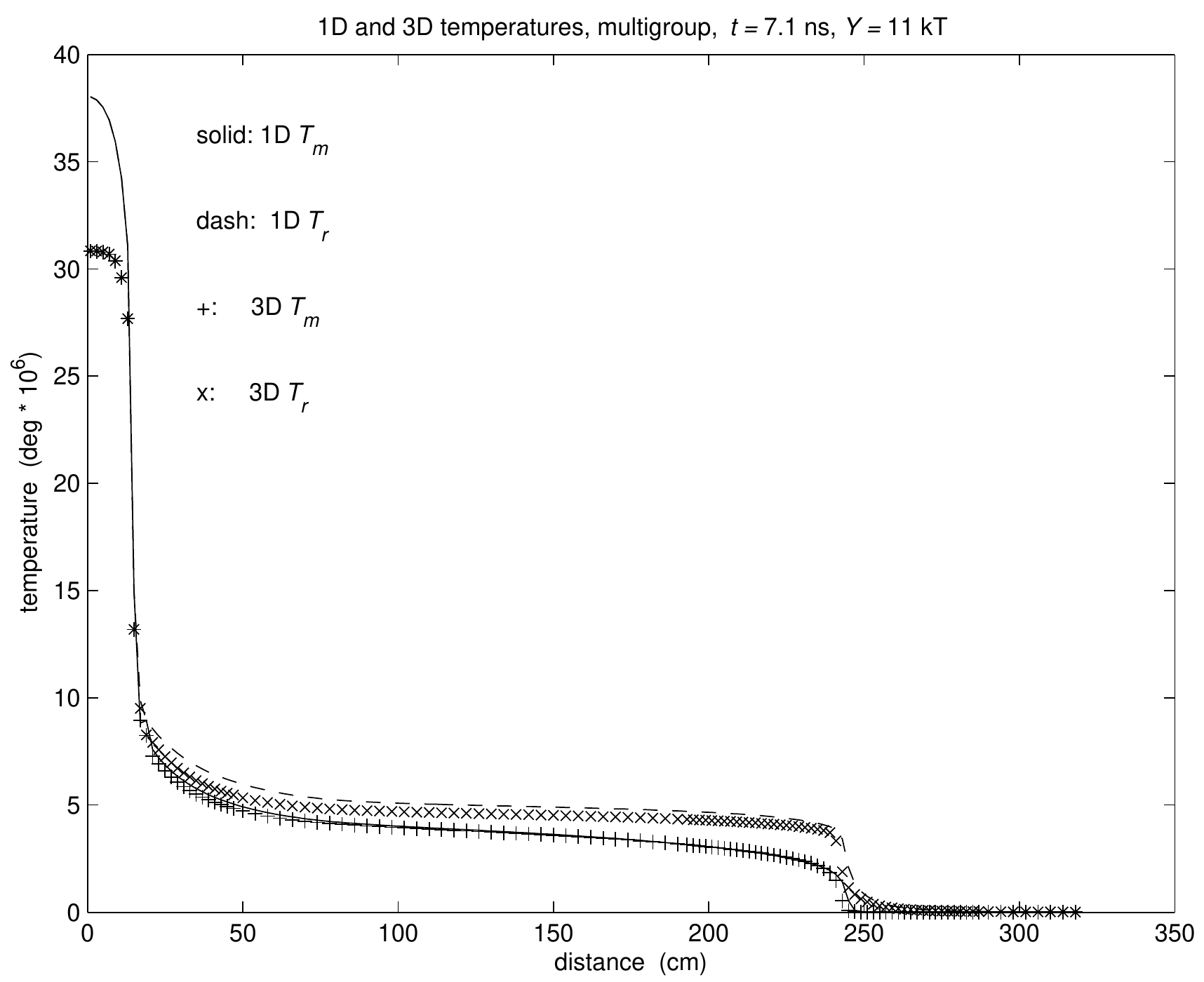}
\caption{Hot sphere problem.
Comparison of $T$ and $T_r$ for
Cartesian and spherical multigroup runs; 
$Y = 11$~kT, $t = 7.1$~ns.}
\label{T1D3D}
\end{center}
\end{figure}

We conclude the section by comparing results of
the 1D spherical and 3D Cartesian versions of our code.
We return to running with full functionality,
i.e., with hydrodynamics, heat conduction, as well as
with two AMR levels.  For the Cartesian simulation, the
Al ``sphere'' is a cube 31~cm per side
(in contrast to the 1D, 31~cm diameter ball.)
The difference in volumes implies that the initial
central, Cartesian temperatures are necessarily smaller
in order to have the same yield.  Figure~\ref{T1D3D}
displays the radial 1D results 
and a $x$ axis lineout of the Cartesian run.
The agreement of the profiles is self-evident. 
 
To summarize, we have simulated 
real-life problems, viz., air bursts with
yields $Y = 11$~kT and 1~MT\@.  We've shown that for low
$Y$, gray and MGD give similar results.  However,
for large $Y$, they differ for early times
when the dynamics are dominated by radiation.  
Our high yield MGD simulation contradicts results of Brode \cite{Bro}.
However, Brode's pioneering simulations
were done many years ago when the relatively limited
computational resources precluded using
sophisticated modules such as MGD.



\section{Conclusion/Summary}
\label{conclusion}

We have described a numerical scheme to solve the 
radiation multigroup diffusion equations.  The scheme is
implemented in a radiation-hydrodynamic code with the
patch-based AMR methodology, originally proposed by 
Berger and Oliger \cite{BeOl} for hyperbolic partial
differential equations.  Our scheme consists of two parts.
The first, described in Sections~\ref{levelsolve}
and \ref{MGanlsys}, is applied on a {\em level\/} of the
AMR grid layout and may be adapted to any code.
This part consists of adding $\ptc$ to the
``fully-implicit'' iterative scheme of Axelrod et al \cite{AxDuRh}.
$\ptc$ brings an extra degree of robustness and enhances
convergence of the Axelrod scheme.  We have developed
lemmas that determine the minimum magnitude for the
$\ptc$ parameter $\tau$ to ensure that the iterations
converge and the result is physically meaningful.
The appropriate magnitude depends on the problem.

Our implementation of $\ptc$ is not optimal---at least
for our AMR code architecture.  In our code,
for each AMR level, we compute a {\em single\/} scalar 
parameter $\tau$.  However, the levels consist of a collection
of grids (rectangles in 2D) that need not be connected.
If the grids are not connected, they form 
independent problems.  Hence, it would be more efficient
to use different $\tau$ for disconnected grids.

The second part of our scheme,  the
sync-solve (SS), addresses a specific
need of our code, viz., the requirement of having an
energy-conserving result on the composite grid of multiple
AMR levels.  For the multigroup equations, this part
reduces to a coupled system of elliptic
equations on the unstructured grid combining all levels.
Since the SS is intended to be a small correction to the
result of the level solves, we adapted the
key element of the ``partial temperature'' scheme of Lund
and Wilson \cite{LunWil}.  This allowed reducing the
multigroup SS to a collection of scalar SS's.  We were then
able to reuse existing software.

This paper included simulations of three problems.
The first two are idealized tests of only the multigroup
module.
The third is a ``real'' problem, which uses the full capability
of the code: AMR, multiple materials, etc. 
The first problem was 
chosen because of its non-triviality and the availability
of analytic results with which to compare.  We obtained
excellent agreement and verified the convergence properties
of the scheme.  The second problem illustrated the benefits
brought by $\ptc$.  We compared the conventional scheme
of Axelrod et al \cite{AxDuRh} with our $\ptc$-modified version.
For hard problems, $\ptc$ either decreased run times or
ensured convergence in regimes where the conventional
scheme diverged.  The third problem showed that our 
multigroup module has been fully integrated into the code
and has already extended the scientific frontier.
For a high yield air burst at STP, we found that gray diffusion 
gives an incorrect result during the radiation-dominated
regime because gray fails to capture the frequency-dependent
effects of the air opacity.

\appendix
\section{Revised S\&B table}
\label{table}

\begin{center}
\begin{tabular}{|lllll|} \hline
$x$           & $T$            & $E_r$   & $\epsilon(T)$ & $\epsilon(E_r)$ \\ \hline
0.0000000E+00 &  9.9373253E-01 & 5.6401674E-03 & 5.4E-09 & 5.9E-11 \\
2.0000000E-01 &  9.9339523E-01 & 5.5646351E-03 & 1.8E-08 & 7.0E-11 \\
4.0000000E-01 &  9.8969664E-01 & 5.1047352E-03 & 6.0E-09 & 6.2E-11 \\
4.6000000E-01 &  9.8060848E-01 & 4.5542134E-03 & 9.8E-09 & 6.4E-11 \\
4.7000000E-01 &  9.7609654E-01 & 4.3744933E-03 & 1.3E-08 & 6.9E-11 \\
4.8000000E-01 &  9.6819424E-01 & 4.1294850E-03 & 8.2E-09 & 6.3E-11 \\
4.9000000E-01 &  9.5044751E-01 & 3.7570008E-03 & 6.7E-09 & 6.3E-11 \\
5.0000000E-01 &  4.9704000E-01 & 2.9096931E-03 & 7.7E-09 & 2.8E-11 \\
5.1000000E-01 &  4.3632445E-02 & 2.0623647E-03 & 1.2E-08 & 6.3E-11 \\
5.2000000E-01 &  2.5885608E-02 & 1.6898183E-03 & 1.3E-08 & 6.3E-11 \\
5.3000000E-01 &  1.7983134E-02 & 1.4447063E-03 & 1.8E-08 & 7.0E-11 \\
5.4000000E-01 &  1.3470947E-02 & 1.2648409E-03 & 1.5E-08 & 6.5E-11 \\
6.0000000E-01 &  4.3797848E-03 & 7.1255738E-04 & 1.1E-08 & 6.4E-11 \\
8.0000000E-01 &  6.4654865E-04 & 2.3412650E-04 & 2.3E-08 & 6.8E-11 \\
1.0000000E+00 &  1.9181546E-04 & 1.0934921E-04 & 1.0E-08 & 6.1E-11 \\ \hline
\end{tabular}
\end{center}
Revised S\&B table (Bolstad \cite{Bol}); time $t = 1.0$, $T_f = 0.1$.
Columns 4 and 5 give maximum, absolute error estimates.  Hence, at $x = 0$,
entry $T$ is correct to $\pm$5.4E-09, i.e., has 8 trustworthy digits. 

\section{Diagonal dominance; large mean-free-paths}
\label{apb}
As noted in the footnote of Section~\ref{diag} (and remarked by a referee),
long mean free paths may lead to diffusion coefficients that overwhelm the 
other matrix terms.  Thus, the estimate for $\gs$, obtained in
Lemma~\ref{lem:L3}, may be insufficient.  The matrix diagonal
 contains three terms of various magnitudes.  The first stems
from the discretization of the $\partial / \partial t$ derivative.
Because we multiply by $\dt$,
the term equals 1.  The second term is due to the coupling coefficient
$a_g$; the term equals $\dt \, c / l_g$, where
$l_g = (\rho \, \gk_g)^{-1}$ is the mean free path.  The third term is the diffusion
coefficient, which after including the time step and discretization
of $\partial^2 / \partial x^2$, is of the form
$\dt \, c \, l_g' / 3 \, h^2$, where $l_g'$ is the flux-limiter-modified
mean free path;
\[
  l_g' = 1/[ (l_g)^{-1} + (3h)^{-1} ( \beta + |\Delta u_g|/u_g ) ] \, ,
\]
where $\beta$ is a small, user-set constant, whose utility will become evident
and $\Delta u_g/u_g$ is a normalized difference of adjoining cell-centered
values.

In the limit $l_g \rightarrow \infty$, the coupling term $a_g$ is negligible.
Hence, we compare the diffusion term with unity.  As $l_g \rightarrow \infty$, 
$l_g'$ no longer depends on $l_g$.  After factoring a
factor of $h$, the diffusion term is of magnitude,
\[
  (\dt \, c / h) \, / \, (\beta + |\Delta u_g|/u_g ) \, .
\]
The quantity $\beta^{-1}$ plays the role of the maximum number of mean
free paths allowed, in units of $h$.  If the gradient of $u_g$ is
not negligible,  $|\Delta u_g|/u_g$ dominates the diffusion term.  If
both $l_g \gg 1$ and $|\Delta u_g|/u_g \ll 1$, $\beta$
dominates.  In that case, we
are left comparing 1 to $\dt \, c / h \, \beta$.  
The parameter $\beta$ is small; we often use $10^{-4}$.  
(However, as shown in appendix~\ref{apc}, $10^{-4}$
is too large.)  Using $c = 3 \cdot 10^{10}$~cm/s
gives a diffusion term of order
\be
  3 \cdot 10^{14} \, \dt /h \, .  \label{dth}
\ee
If this exceeds machine precision, the $\gs$ estimate of
Lemma~\ref{lem:L3} does not guarantee diagonal dominance.
We are now left with problem-specific estimates.  
Clearly, simulations requiring small $h$ or large $\dt$ are
problematic.  Luckily, our envisioned applications yield
reasonable $\dt /h$ ratios.  

Consider two topics, ICF hohlraums and 
simulations of the type described in Section~\ref{hotball}.  For the
former, mesh sizes are rarely less than
0.1 microns, i.e., min($h) = \mathcal{O} (10^{-5})$~cm.
Luckily, in ICF, typical total simulations times are of order of
tens of ns, requiring significantly smaller timesteps.  Using
max($\dt) = \mathcal{O} (10^{-9})$~s, makes \pref{dth} of
order $10^{10}$, which, when compared to unity, is six
orders of magnitude above double precision.

For applications of the type presented in section~\ref{hotball}, while
timesteps vary enormously, so do mesh sizes; hence, the ratio $\dt /h$
remains moderate.    For long-time
simulations requiring $\dt$ exceeding 1~s, it is unlikely
that it is necessary to resolve details less than 100 cm.  Substituting
these values into \pref{dth} leaves $3 \cdot 10^{12}$, which, is
also resolved by double precision, but just barely.

\section{Full physics convergence analysis}
\label{apc}

This section presents a 
spatial convergence analysis of the scheme as it may be used in practice.
Particular attention is devoted to effects of the flux limiter and AMR.
In contrast to what was analyzed at the end of section~\ref{linwin},
here we refine about a moving front.

The exactness of the solution depends upon the magnitudes of $\dt$ and $\dx$.  
To ensure that the time step does not dominate the error, we use a
very conservative value for $\dt$, which is much smaller than what would be used in practice. 
When $\dt$ is small,
$\ptc$ is not needed. 
Furthermore, we find that solutions using the FI and SI schemes are
indistinguishable for our chosen $\dt$.  We obtain the same result by solving
nonlinear problems for each time step (FI) as by linearizing the equations
and solving linear systems (SI).  In order to save computer time,
the simulations in this section
use the SI scheme and do not use $\ptc$.

To address concerns of a referee, we consider a stringent test
and focus attention on the problem
described in section~\ref{hotball}.
An Al sphere of 15.5 cm radius is
suspended in air.  Initially, both sphere and air are at STP;
the radiation field is initially zero. 
We load 
a $Y = 1$ MT source (approximately $4 \cdot 10^{22}$ erg) into
the radiation field only in the region containing Al.
The source is loaded into a Planckian spectrum
over a time interval $t_s = 0.1$ ns.  The interval
is so short that over its duration the main
effects are to raise the radiation field to a high temperature
and to a lesser extent also increase the
matter temperature due to coupling.  At $t = t_s$, most of the
energy is in the radiation field inside the sphere.
The radiation temperature $T_r$ is
largely uniform over the sphere and equals approximately $1.3 \cdot 10^8$
deg, or nearly 12 keV.

To highlight effects of the
flux limiter, we examine the solution at $t = 10^{-7}$~s.
In order to analyze errors due to only our multigroup scheme,
we turn off all other physics, e.g., hydrodynamics.  This yields profiles
that are similar to those obtained with a ``full physics'' simulation
since at $t = 10^{-7}$~s hydro effects should be negligible. 
(Assuming maximum
sound/shock speeds of $\co(10^{7})$~cm/s, the most that hydrodynamics
can do is push the Al/air interface out a few cm while the hot sphere can
radiate out to 3000 cm.)  

\begin{figure}
\begin{center}
\includegraphics[width=\fsize]{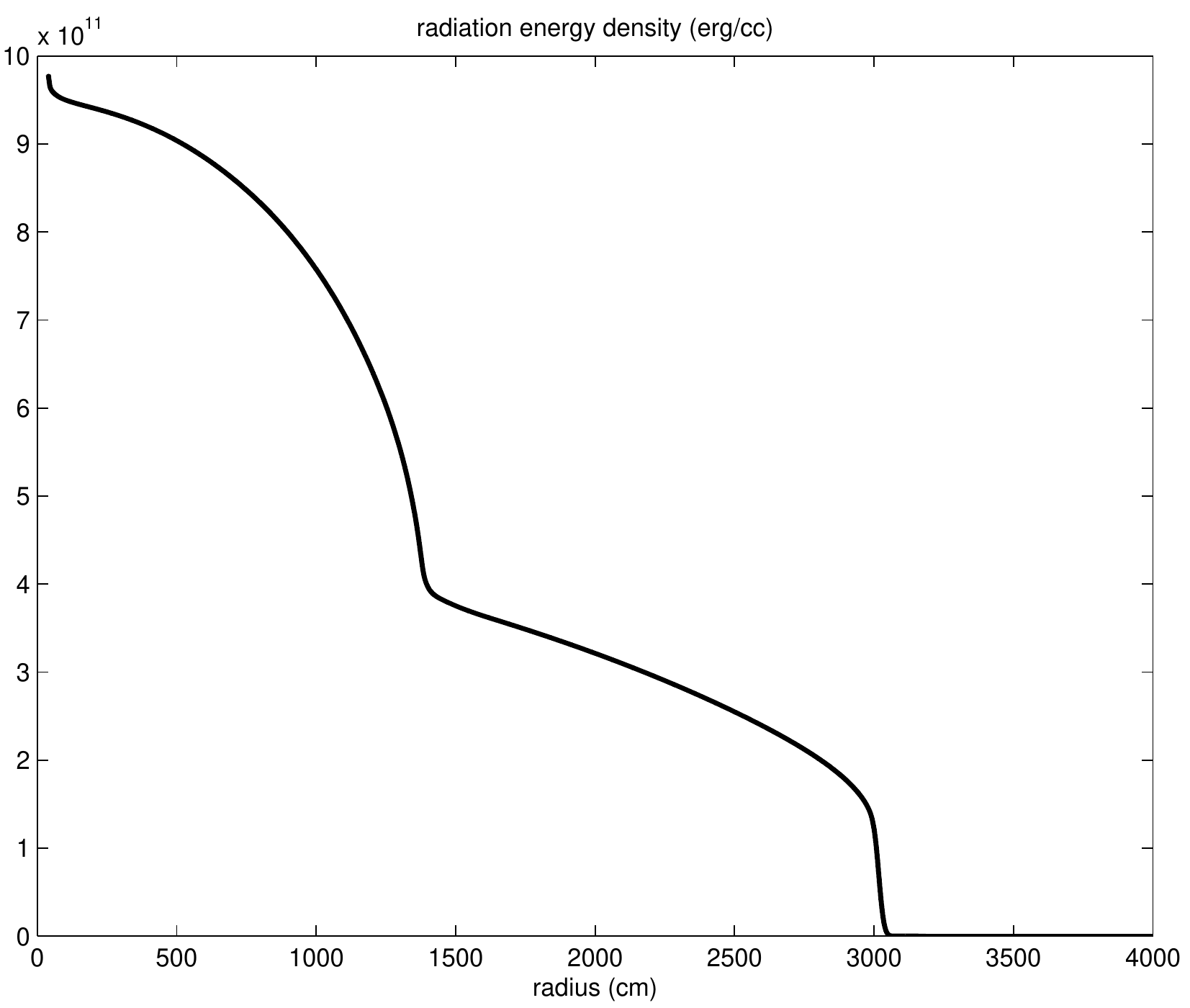} 
\caption{Hot sphere problem. Radiation energy density $E_r$ (erg/cc),
$Y = 1$ MT, $t = 0.1 \,\mu$s,
$\dx = 0.5$~cm.  Air region.}
\label{Er}
\end{center}
\end{figure}

\begin{figure}
\begin{center}
\includegraphics[width=\fsize]{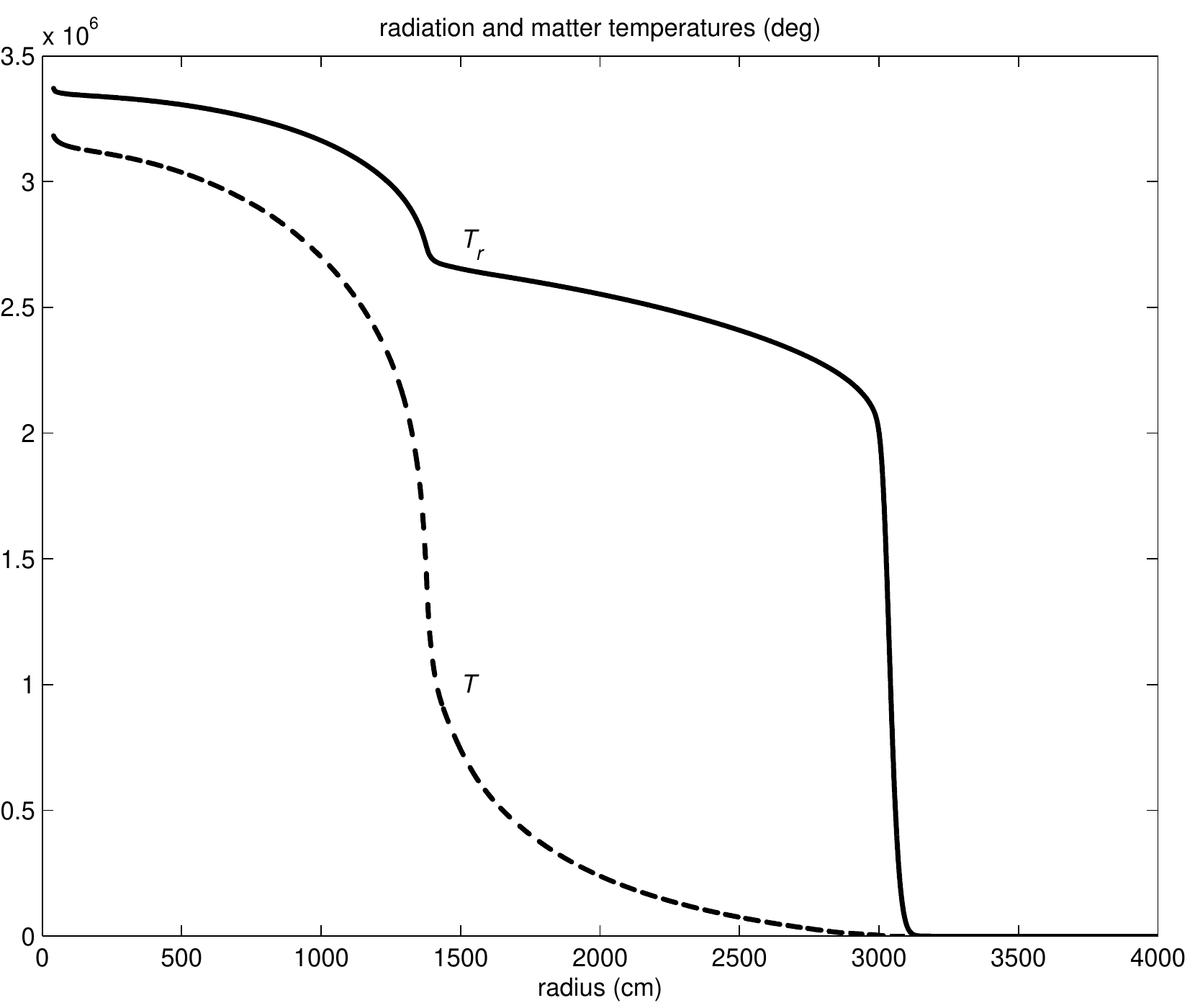} 
\caption{Hot sphere problem. Radiation $T_r$ and matter temperature
$T$ (deg), $Y = 1$ MT, $t = 0.1 \,\mu$s,
$\dx = 0.5$~cm.  Air region.}
\label{Tr_T}
\end{center}
\end{figure}

We examine the total radiation energy density;
Fig.~\ref{Er} displays $Er$ for $r > 40$~cm.
Inside the Al, $E_r$ is much larger than what is shown in Fig.~\ref{Er};
it decays steeply from a central value of $2.6 \cdot 10^{14}$ erg/cc,
to $1.5 \cdot 10^{12}$ at $r = 16$, which designates the air cell adjoining the sphere.
Hence, the radiation temperature decays from a central value of 
$T_r = 14 \cdot 10^{6}$ to $3.8 \cdot 10^{6}$~deg.
The change in slope at 1400 cm is explained by examining
the temperatures $T_r$ and $T$; see Fig.~\ref{Tr_T}.
The distance $r =  1400$ marks the approximate extent of the
fireball.  However, radiation propagates out to $r = 3000$ cm,
then drops sharply; the drop due to the flux limiter. 

We identify three distinct regions in the profiles of Figs.~\ref{Er}
and \ref{Tr_T}.  The innermost, out to $r =  1400$ denotes where
we can expect the {\em diffusion\/} equations to
yield an accurate representation of the physics.
There, the domain is largely optically thick, as evidenced
by the close agreement of $T$ and $T_r$.  The region $1400 < r < 2900$
denotes an optically thin region, where the diffusion approximation
is expected to fail.  Lastly, in the region $r > 2900$ the solution depends entirely  on
a {\em kludge:} the flux limiter.  However, although at this time
the limiter is dominant only near $r = 3000$, it has
affected the entire solution because the propagation of the front is
governed by the limiter and hence all cells out to the
present position of the front have been traversed by the
leading edge of the wave. 

For the purposes of the convergence study, we define the results displayed in
Figs.~\ref{Er} and \ref{Tr_T} as the ``converged'' solution. 
We obtain
it using a uniform grid $\dx = 0.5$ cm and an initial $\dt =10^{-16}$~s.
The timestep increases by 5\% each cycle but is not allowed
to exceed  $2 \cdot 10^{-12}$~s.  We take 50,183 steps to reach the
final time. 
The discretization yields a light-speed
Courant number $ C_c \doteq c \dt / \dx = 0.12$.  Although the value
may seem overly cautious, it is still too large.
A transport calculation would preclude any signal from propagating beyond
$r = 3016$ cm.  However, our finest-grid diffusion result yields
$T_r = 1.8 \cdot 10^6 $ and 12,000  deg at $r = 3016$ and 3116 cm, respectively.
Although the enhanced diffusion of our result may be due to our
choice of a limiter, and as analyzed by Morel \cite{Mor2} and Olson et al
\cite{OlAu} there are other limiter choices,
all limiters reduce to discretizing the equation $u_t = c u_x$.

We time-lag the limiter for two reasons.
(1) Flux-limiting is a kludge. Thus, a time-advanced limiter is not only more complicated to implement but it does not yield a more accurate solution.
(2) When a front propagates into cold
material, a time-lagged limiter puts a front slightly behind 
where a time-advanced limiter would place it.\footnote{The result may
be seen by comparing two face-centered, flux-limited diffusion coefficients
of the form $u/|\partial_x u|$
and discretized as $(h/2) (u_0 + u_1)/|u_0 - u_1|$.
Assume the front propagates into cell $0$.  Let the time-lagged
$u_0 = 0$ and the time advanced values be $u_0'$ and $u_1'$ with
$u_0' \ll u_1'$.  After dividing by $h/2$,
the time-lagged and time-advanced coefficients equal 
1 and $(1 + \gvep)/(1 - \gvep)$, respectively,
where $\gvep = u_0' / u_1'$.  Thus, the time-lagged diffusion
is smaller and the front does not propagate as far.}  
So, since the raison d'etre of a limiter is to retard the flow, we
time-lag.

Before describing the convergence study, we take up two
topics.  The first is a truncation error analysis of a
diffusion equation with a time-lagged flux limiter.
If the limiter is dominant, the face-centered diffusion
coefficient is
$D_{i+1/2} = [(u_{i+1} + u_i)/2]/[|u_{i+1} - u_i|/\dx]$,
where $i$ denotes the cell index.
After inserting this expression into the equation
discretized with backward Euler we obtain,
\be
   u_i' - u_i = (C_c/2) \, 
   [ \, G_{i+1/2} \, (u_{i+1} + u_i) - G_{i-1/2} \, (u_{i} + u_{i-1})
   \, ] \, , \label{ApC_eq1}
\ee
where $G_{i+1/2} = (u_{i+1}' - u_i')/|u_{i+1} - u_i|$
and primes denote the time-advanced variable.
We ignore the absolute value operator
since it only serves to enforce flow down the gradient.
Since $G$ is of the form $f(t + \dt)/f(t)$,
$G_{i \pm 1/2} = 1 + \dt \, F$, where $F$ has units of
inverse time and its leading term is $f'(t)/f(t)$.
Expanding the LS of \pref{ApC_eq1} about the time-retarded value
yields, $\dt \, \partial_t u + \co(\dt)^2$, while inserting
the expansions of $G_{i \pm 1/2}$ reduces the RS of \pref{ApC_eq1} to
\[
   (C_c/2) \, 
   [ \, (u_{i+1} - u_{i-1}) + \dt \, U_t
   \, ] \, . 
\]
The term $U_t$ is of the form $(\partial_t f /f) \, \dx \, \partial_x u$
and has units of $u/t$.
The difference $(u_{i+1} - u_{i-1})$ yields
$2 \dx \, \partial_x u + \co(\dx^3)$.
After simplifying, we obtain the truncation error of \pref{ApC_eq1},
\[
  \partial_t u + \co(\dt) = 
  c \, \partial_x u + \co(\dx^2) + C_c \, U_t/2 \, .
\]
The $C_c$ term is important.
It shows that even when both $\dt$ and $\dx$ are small, the
discretization has an additional error proportional to the
light-speed Courant number. 

The second topic is related to the parameter $\gb$ introduced
at the end of section~\ref{levelsolve}. 
Appendix~\ref{apb} shows that for large mean free paths
the diffusion coefficient may depend solely on the
sum $\beta/\dx + |\nabla u|/u$.  Consider Fig.~\ref{Er}
and the domain $1500 < r < 2500$.  From values at
$r = 1500$, 2000, 3000, we estimate the
average of $|\nabla E_r|/E_r$ to be $3.7 \cdot 10^{-4}$.
Thus, if a particular $\gb$ is deemed
sufficiently small for some coarse mesh width,
as the mesh is refined, the ratio $\gb / \dx$ may overwhelm
the flux limiter.  The statement has implications for both
a conventional uniform-grid convergence study as well as
one with AMR, since we use the same $\gb$ on all levels.
Unless stated otherwise, for all runs discussed in this section
$\gb/\dx = 10^{-6}$.

We now describe the single-level convergence study.
We make five runs with successively finer grids,
$\dx = 8$, 4, 2, 1, and for the ``converged'' result, $\dx = 0.5$~cm.
All runs have the same initial $\dt$ time history.  
However, 
$\max(\dt)$ depends on the size of $\dx$ in order to maintain
$\max(C_c)$ at 0.12.  Hence,
$\max(\dt)$ varies from $3.2 \cdot 10^{-11}$ to
$2 \cdot 10^{-12}$ between the coarsest and finest runs.

Our code computes
cell-averaged quantities, e.g., $E_r$ (erg/cc).
On the coarsest grid, we obtain $500$ values
on cells centered at 4, 12, 20, $\ldots$, cm.  
In order to have an equitable comparison, for each
run, we also determine 500 average values by post-processing.
In order to conserve energy, we use volume averaging.

The error is computed as follows.
Let $E_{i, \,\dx}$ denote the averaged energy
density in the $i$th coarse cell for a run with
mesh size $\dx$.  By defining the relative error,
\be
  e_{\dx} = \left. \left[ \sum_{i = 1}^{500}
    (E_{i, \, \dx} - E_{i,\,0.5} )^2 \, \right]^{1/2} \, 
            \right/ \left[ \sum_{i = 1}^{500}
    (E_{i,\,0.5} )^2 \, \right]^{1/2} \, , \label{errdef}
\ee
we obtain
\[
  [ \, e_8, \,  e_4, \, e_2, \, e_1] =
  [ \, 0.4820, \, 0.2436, \, 0.06969, \, 0.03475 ]  \, .  
\]
The ratio of successive errors,
\[
    [ \, (e_8/e_4),  \, (e_4/e_2), \, (e_2/e_1)] =
  [ \, 1.98, \, 3.50, \, 2.01 ] \, .
\]
Since the ratios are approximately two, our results
suggest first (rather than second) order convergence.

Lastly, we compare results of a run using AMR with an ``equivalent''
run that uses a uniform grid.  For the AMR run, we use a base L0 grid with
$\dx = 8$~cm.  We use two refinement levels, each refines by a factor of two.
The run begins with $\dt = 10^{-16}$, and we increase $\dt$ as before
until reaching $\max(\dt) = 3.2 \cdot 10^{-11}$.
Because we refine in both space and time, $C_c = 0.12$ on all levels.
The parameter $\gb = 8 \cdot 10^{-6}$; hence,
$\gb/\dx$ varies from $10^{-6}$ to $4 \cdot 10^{-6}$
between the coarsest and finest levels.
The refined levels adapt to the Al/air interface
(at $r = 15.5$~cm) and around the position of
$\max[ |\nabla(E_r)|/E_r]$.
At the end of the run, the grid layout is: 
$0 < x < 96$ and $2944 < x < 3392$ for the L1 level ($\dx = 4$)
while:
$0 < x < 48$ and $2976 < x < 3360$ for L2 ($\dx = 2$).
We compare errors of the AMR run as above, by forming
cell averages of the 500 cells centered at 4, 12, 20, $\ldots$, cm.
The error on the AMR run is
\[
  e_{AMR} = 0.06963  \, .
\]
Since for the AMR run, the finest level $\dx = 2$~cm, we
compare $e_{AMR}$ with the error of a uniform-grid run
where $\dx = 2$~cm, i.e., with $e_2 = 0.06969$.
To three significant digits, the errors
are essentially equal.
The uniform-grid run uses 2000 cells, while 
at the end of the simulation, the AMR run has 676 cells.

To summarize, we find that when the solution depends on the
flux limiter, spatial convergence reduces to
being first, instead of second order.  The effect reminds us
of what happens to hydrodynamic schemes in
the presence of shocks: in smooth parts of the flow, the
solution may be second order convergent, but in regions
traversed by shocks, the scheme reverts to first order.
In closing, we note an additional source of error, which
we did not quantify.  The simulations of this section
use real materials whose properties (internal energies,
opacities, etc.) are given by tables.  Table lookups
have errors that depend on the schemes used to interpolate
between table data.

\vspace{.25in}
\large
\noindent
{\bf Acknowledgment }
\normalsize
We thank Dr.\ J.\ Bolstad (LLNL) for computing the revised data displayed
in the Appendix and for a careful reading of the manuscript.
We are grateful to Dr.\ M.\ Clover (SAIC) for many fruitful discussions.
We also thank the referees for a careful and thoughtful review of
our original manuscript.  The paper is very much improved due to
their suggestions.  Unfortunately, they must remain anonymous.
  
\end{document}